\documentclass[11pt,doublespace]{article} 
\usepackage{graphics} 
\usepackage[pdftex]{graphicx}
\usepackage [cmex10]{amsmath} 
\usepackage{amssymb}  

      \newtheorem{theorem}{Theorem}
      \newtheorem{lemma}{Lemma}

      \newtheorem{definition}{Definition}
      
\begin{document}

\title{Distributed Joint Source-Channel Coding on a Multiple Access Channel with Side Information}

\author{R Rajesh and Vinod Sharma\\
 Dept. of Electrical Communication Engg. \\
Indian Institute of Science \\ Bangalore, India\\Email: rajesh@pal.ece.iisc.ernet.in,~vinod@ece.iisc.ernet.in}

\maketitle
\thispagestyle{empty}
\pagestyle{empty}

\begin{abstract}
We consider the problem of transmission of several distributed sources over a multiple access channel (MAC) with side information at the sources and the decoder. Source-channel separation does not hold for this channel. Sufficient conditions are provided for transmission of sources with a given distortion. The source and/or the channel could have continuous alphabets (thus Gaussian sources and Gaussian MACs are special cases). Various previous results are obtained as special cases. We also provide several good joint source-channel coding schemes for a discrete/continuous source and discrete/continuous alphabet channel. Channels with feedback and fading are also considered.
\end{abstract}

\section{Introduction}\label{sec1.1}
In this report we consider the transmission of various sources over a multiple access channel (MAC). We survey the result available when the system may have side information at the sources and/or at the decoder. We also consider a MAC with feedback or when the channel experiences time varying fading.

This system does not satisfy source-channel separation (\cite{Cover80multiple}). Thus for optimum transmission one needs to consider joint source-channel coding. Thus we will provide several good joint source-channel coding schemes. Although this topic has been studied for last several decades, one recent motivation is the problem of estimating a random field via sensor networks. Sensor nodes have limited computational and storage capabilities and very limited energy \cite{akylidiz02survey}. These sensor nodes need to transmit their observations to a fusion center which uses this data to estimate the sensed random field. Since transmission is very energy intensive, it is important to minimize it.

The proximity of the sensing nodes to each other induces high correlations between the observations of adjacent sensors. One can exploit these correlations to compress the transmitted data significantly. Furthermore, some of the nodes can be more powerful and can act as cluster heads (\cite{Baek04minimizing}). Neighboring nodes can first transmit their data to a cluster head which can further compress information before transmission to the fusion center. The transmission of data from sensor nodes to their cluster-head is usually through a MAC. At the fusion center the underlying physical process is estimated. The main trade-off possible is between the rates at which the sensors send their observations and the distortion incurred in the estimation at the fusion center. The availability of side information at the encoders and/or the decoder can reduce the rate of transmission (\cite{wyner},\cite{Gastpar}).

The above considerations open up new interesting problems in multi-user information theory and the quest for finding the optimal performance for various models of sources, channels and side information have made this an active area of research. The optimal solution is unknown except in a few simple cases. In this report a joint source channel coding approach is discussed under various assumptions on side information and distortion criteria. Sufficient conditions for transmission of discrete/continuous alphabet sources over a discrete/continuous  alphabet MAC are given. These results generalize the previous results available on this problem.

The report is organized as follows. Section~\ref{sec1.2} provides the background  and surveys the related literature. Transmission of distributed sources over a MAC with side information is considered in section~\ref{sec1.3}. The sources and the channel alphabets can be continuous or discrete. Several previous results are recovered as special cases in section~\ref{sec1.4}. Section~\ref{sec1.5} considers the important case of transmission of discrete correlated sources over a Gaussian MAC (GMAC) and presents  a  new coding scheme. Section~\ref{sec1.6} discusses several joint source-channel coding schemes for transmission of Gaussian sources over a GMAC and compares their performance. It also suggests coding schemes for general continuous sources over a GMAC. Transmission of correlated sources over orthogonal channels is considered in section~\ref{sec1.7}. Section~\ref{sec1.8} discusses a MAC with feedback. A MAC with multi path fading is addressed in section~\ref{sec1.9}. Section~\ref{sec1.10} provides practical schemes for joint source-channel coding. Section~\ref{sec1.11} gives the directions for future research and section~\ref{sec1.12} concludes the report.

\section{Background}\label{sec1.2}
In the following we survey the related literature. Ahlswede (\cite{ash}) and Liao (\cite{liao}) obtained the capacity region of a discrete memoryless MAC with independent inputs. Cover, El Gamal and Salehi in \cite{Cover80multiple} made further significant progress by providing sufficient conditions for transmitting losslessly correlated observations over a MAC. They proposed a `correlation preserving' scheme for transmitting the sources. This mapping is extended to a more general system with several principle sources and several side information sources subject to cross observations at the encoders in \cite{han}. However single letter characterization of the capacity region is still unknown. Indeed Duek \cite{Duek81note} proved that the conditions given in \cite{Cover80multiple} are only sufficient and may not be necessary. In  \cite{kang} a single letter upper bound for the problem is obtained. It is also shown in \cite{Cover80multiple} that the source-channel separation does not hold in this case.  The authors in  \cite{Medrad06seperation} obtain a condition for separation to hold in a multiple access channel. 

The capacity region for distributed lossless source coding problem is given in the classic paper by Slepian and Wolf (\cite{slepian}). Cover (\cite{cov}) extended Slepian-Wolf results to an arbitrary number of discrete, ergodic sources using a technique called `random binning'. Other related papers on this problem are \cite{serv},\cite{han}.

Inspired by Slepian-Wolf results, Wyner and Ziv \cite{wyner} obtained the rate distortion function for source coding with side information at the decoder. Unlike for the lossless case, it is shown that the knowledge of the side information at the encoders in addition to the decoder, permits the transmission  at a lower  rate. The latter result when encoder and decoder have side information was first obtained by Gray and is known as conditional rate distortion function (See \cite{berger}). Related work on side information coding is \cite{wornell,pr,draper}. The lossy version of Slepian-Wolf problem is called multi-terminal source coding problem and despite numerous attempts (e.g., \cite{by},\cite{Oohama97Gaussian}) the exact rate region is not known except for a few special cases. First major advancement was in  Berger and Tung (\cite{berger}) where an inner and an outer bound on the rate distortion region was obtained. Lossy coding of continuous sources at the  high resolution limit is given in \cite{zamir} where an explicit single-letter bound is obtained. Gastpar (\cite{gasp}) derived an inner and an outer bound  with side information and proved the tightness of his bounds  when the sources are conditionally independent given the side information. The authors in \cite{ncc} obtain inner and outer bounds on the rate region with side information at the encoders and the decoder. References~\cite{Varsheneya06distributed},\cite{Rajesh07gaussian} extend the result in \cite{ncc} by requiring the encoders to communicate over a MAC, i.e., they obtain sufficient conditions for transmission of correlated sources over a MAC with  given distortion constraints. In \cite{ong} achievable rate region for a MAC with correlated sources and feedback is given.

The distributed Gaussian source coding problem is discussed in \cite{Oohama97Gaussian},\cite{Wagner05rate}. Exact rate region is provided in \cite{Wagner05rate}. The capacity of a Gaussian MAC (GMAC) with feedback is given in \cite{Ozarow84capacity}. In \cite{Lapidoth01sending} a necessary and two sufficient conditions for transmitting a jointly Gaussian source over a GMAC are provided. It is shown that the amplify and forward scheme is optimal below  a certain SNR determined by source correlations. The performance comparison of the schemes given in \cite{Lapidoth01sending} with a Separation based scheme is given in \cite{Rajesh07allerton}. GMAC under received power constraints is studied in \cite{Gastpar04gaussian} and it is shown that the source-channel separation holds in this case. 

In~\cite{Gastpar03source} the authors discuss a joint source channel coding scheme over a MAC and show the scaling behavior for the Gaussian channel. A Gaussian sensor network in distributed and collaborative setting is studied in \cite{Iswar05rate}. The authors show that it is better to compress the local estimates than to compress the raw data. The scaling laws for a many-to-one data-gathering channel are discussed in \cite{elgmal}. It is shown that the transport capacity of the network scales as $\mathcal{O}(logN)$ when the number of sensors $N$ grows to infinity and the total average power remains fixed. The scaling laws for the problem without side information are discussed in \cite{Gastpar05power} and it is shown that  separating source coding from channel coding may require exponential growth, as a function of number of sensors, in communication bandwidth. A lower bound on best achievable distortion as a function of the number of sensors, total transmit power, the  degrees of freedom of the underlying process and the spatio-temporal communication bandwidth is given.

 The joint source-channel coding problem also bears relationship to the CEO problem \cite{ceo}. In this problem, multiple encoders observe different, noisy versions of a single information source and communicate it to a single decoder called the CEO which is required to reconstruct the source within a certain distortion. The Gaussian version of the CEO problem is studied in \cite{Oohama98rate}. 

When Time Division Multiple Access (TDMA), Code Division Multiple Access (CDMA) or Frequency Division Multiple Access (FDMA) are used then a MAC becomes a system of orthogonal channels. These protocols, although suboptimal are frequently used in practice and hence have been extensively studied (\cite{Cover04elements},\cite{pbook}). Lossless transmission of correlated sources over orthogonal channels is addressed in \cite{barros}. The authors prove that the source-channel separation holds for this system. They also obtain the exact rate region. Reference~\cite{xiong} extends these results to the lossy case and shows that separation holds for the lossy case too. Distributed scalar quantizers  were designed for correlated Gaussian sources and independent Gaussian channels in \cite{karlson07isit}. 

The information-theoretic and communication features of a fading MAC are given in an excellent survey paper \cite{proakis}. A survey of practical schems for distributed source coding for sensor networks is given in \cite{dsc}. Practical schemes for distributed source coding are also provided in \cite{discus},\cite{eldpc}.     


\section{Transmission of correlated sources over a MAC}\label{sec1.3}
\begin{figure}[h]
\centering
\includegraphics [width=8.5cm,height=3.6cm] {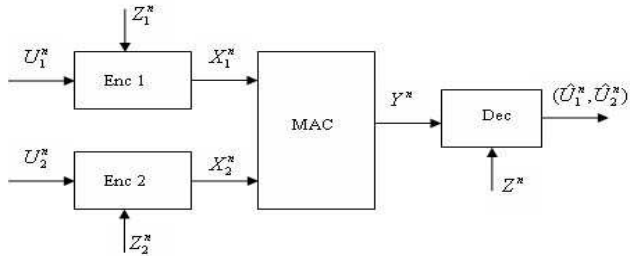}
\caption{Transmission of correlated sources over a MAC with side information}
\label{fig1.1}
\end{figure}

In this section we consider the transmission of memoryless dependent sources, through a memoryless multiple access channel  (Fig.~\ref{fig1.1}). The sources and/or the channel input/output alphabets can be discrete or continuous. Furthermore, side information may be available at the encoders and the decoder. Thus our system is very general and covers many systems studied over the years as special cases.
We consider two sources $(U_1,U_2)$ and side information random variables $Z_1, Z_2, Z$ with a known joint distribution $F(u_1, u_2, z_1, z_2, z)$. Side information $Z_i$ is available to encoder $i,~i\in \{1, 2\}$ and the decoder has side information $Z$. The random vector sequence $\{(U_{1n}, U_{2n}, Z_{1n}, Z_{2n}, Z_n), n\geq1\}$ formed from the source outputs and the side information with distribution $F$ is independent identically distributed ({\it{iid}}) in time. The sources transmit their codewords $X_i$'s to a single decoder through a memoryless multiple access channel. The channel output $Y$ has distribution $p(y|x_1,x_2)$ if $x_1$ and $x_2$ are transmitted at that time.  The decoder receives $Y$ and also has access to the side information $Z$. The encoders at the two users do not communicate with each other except via the side information. It uses $Y$ and $Z$ to estimate the sensor observations $U_i$   as $ \hat{U_i},~i\in \{1, 2\}$. It is of interest to find encoders and a decoder such that  $\{U_{1n},U_{2n},n\geq1\}$ can be transmitted over the given MAC with $E[d_1(U_1,\hat{U_1})]\leq D_1$ and  $E[d_2(U_2,\hat{U_2})]\leq D_2$
 where $d_i$  are non-negative distortion measures and $D_i$ are the given distortion constraints. If the distortion measures are unbounded we assume that $u_i^*,~i =1, 2$ exist such that $E[d_i(U_i,u_i^*)] < \infty,~i = 1, 2$.  Source channel separation does not hold in this case.

For discrete sources a common distortion measure is Hamming distance
\begin{eqnarray*}
d(x,x') &=&1~\text{if}~x \ne x',\\
d(x,x')&=&0~\text{if}~x = x'.
\end{eqnarray*}
For continuous alphabet sources the most common distortion measure is $d(x,x')=(x-x')^2$.

We will denote $\{U_{ij},j=1,2,...,n\}$  by $U_i^n, i=1,2$.

\begin{definition}
The source $(U_1^n ,U_2^n)$ can be transmitted over the multiple access channel with distortions ${\bf{D}}{\buildrel\Delta \over=}(D_1,D_2)$ if for any $\epsilon  > 0 $ there is an $n_0$ such that for all $n > n_0$  there exist encoders $f_{E,i}^n: \mathcal{U}_i^n \times \mathcal{Z}_i^n \rightarrow \mathcal{X}_i^n , i \in \{1,2\}$ and a decoder ${{f}_{D}^n}: \mathcal{Y}^n \times \mathcal{Z}^n \rightarrow (\mathcal{\hat{U}}_1^n , \mathcal{\hat{U}}_2^n) $   such that $ \frac{1}{n}E\left[\sum_{j=1}^nd (U_{ij},\hat{U}_{ij})\right]\leq D_i+ \epsilon,~i\in\{1,2\}$ where $(\hat{U}_1^n,\hat{U}_2^n )= f_D(Y^n,Z^n)$,  $\mathcal{U}_i,~\mathcal{Z}_i,~\mathcal{Z},~\mathcal{X}_i,~\mathcal{Y},~\hat{\mathcal{U}_i} $  are the sets in which $ U_i,~Z_i,~Z,~X_i,~Y,~\hat{U}_i$ take values.
\end{definition}    

We denote the joint distribution of $(U_1, U_2)$ by $p(u_1,u_2)$ and  let $p(y|x_1,x_2)$ be the transition probabilities of the MAC. Since the MAC is memoryless, $p(y^n|x_1^n,x_2^n)=\prod_{j=1}^n p(y_j|x_{1j},x_{2j})$. $X\leftrightarrow Y \leftrightarrow Z$   will indicate that $\{X, Y, Z\}$ form a Markov chain.

Now we state the main Theorem. 
\begin{theorem}
\label{theorem1}
	 A source  can be transmitted over the multiple access channel with distortions $(D_1, D_2)$ if there exist random variables $(W_1, W_2,  X_1,  X_2)$ such  that
\begin{eqnarray*}
1.~ p(u_1,u_2,z_1,z_2,z,w_1,w_2,x_1,x_2,y) = 
      p(u_1,u_2,z_1,z_2,z)
 p(w_1|u_1,z_1) \\ p(w_2|u_2,z_2)  
      p(x_1|w_1)p(x_2|w_2)p(y|x_1,x_2)  
\end{eqnarray*}
and \\

 $2$. there exists a function $f_D: \mathcal{W}_1 \times \mathcal{W}_2 \times \mathcal{Z} \rightarrow (\hat{\mathcal{U}}_1 \times  \hat{\mathcal{U}}_2) $  such that $ E[d(U_{i},\hat{U}_{i})]\leq D_i,~i=1,2$, where $(\hat{U}_1,\hat{U}_2)=f_D(W_1,W_2,Z)$ and the constraints
\begin{eqnarray}
I (U_1,Z_1; W_1 | W_2,Z)  &<&  I (X_1; Y | X_2, W_2, Z),\nonumber\\
I (U_2,Z_2; W_2 | W_1,Z) &<& I (X_2; Y | X_1, W_1, Z),\label{eqn1.1}\\	   		    
I (U_1,U_2,Z_1,Z_2 ; W_1, W_2 |Z) &<& I (X_1, X_2; Y |Z)\nonumber
\end{eqnarray}
are satisfied where $\mathcal{W}_i$  are the sets in which $ W_i$ take values.
\end{theorem}

In Theorem~\ref{theorem1} the encoding scheme involves distributed quantization $(W_1,W_2)$ of the sources $(U_1,U_2)$ and the side information $Z_1,Z_2$ followed by correlation preserving mapping to the channel codewords $(X_1,X_2)$. The decoding approach involves first decoding $(W_1,W_2)$ and then obtaining estimate $(\hat{U}_1,\hat{U}_2)$ as a function of $(W_1,W_2)$ and the decoder side information $Z$. The proof of the theorem is given in Appendix A.

If the channel alphabets are continuous (e.g., GMAC) then in addition to the conditions in Theorem~\ref{theorem1} certain power constraints $E[X_i^2] \le P_i,~i=1,2$ are also needed.

For the discrete sources to recover the results with lossless transmission one can use Hamming distance as the distortion measure.

If the source-channel separation holds then one can talk about the capacity region of the channel. For example, when there is no side information $Z_1,Z_2,Z$ and the sources are independent then we obtain the rate region
\begin{eqnarray}
R_1 \leq I(X_1;Y|X_2),~
R_2 \leq I(X_2;Y|X_1),\nonumber\label{ne}\\
R_1+ R_2 \leq I(X_1,X_2;Y).
\end{eqnarray}
This is the well known rate region of a  MAC (\cite{Cover04elements}). Other special cases will be provided in Sec.~\ref{sec1.4}.

In Theorem~\ref{theorem1} it is possible to include other distortion constraints. For example, in addition to the bounds on $E[d(U_i,\hat{U}_i)]$ one may want a bound on the joint distortion $E[d((U_1,U_2),(\hat{U}_1,\hat{U}_2))]$. Then the only modification needed in the statement of the above theorem is to include this also as a condition in defining $f_D$.

If we only want to estimate a function $g(U_1,U_2)$ at the decoder and not $(U_1,U_2)$ themselves, then again one can use the techniques in proof of Theorem~\ref{theorem1} to obtain sufficient conditions. Depending upon $g$, the conditions needed may be weaker than those needed in \eqref{eqn1.1}

The main problem in using Theorem~\ref{theorem1} is in obtaining good source-channel coding schemes providing $(W_1,W_2,X_1,X_2)$ which satisfy the conditions in the theorem for a given source $(U_1,U_2)$ and channel. A substantial part of this report will be devoted to this problem.

\subsection{ Extension to multiple sources} 

The above results can be generalized to the multiple $(\geq2)$ source   case. Let $\mathcal{S}={1,2,...,M}$ be the set of sources with joint distribution $p(u_1,u_2...u_M)$.
	
\begin{theorem}
Sources $(U_i^n ,i\in \mathcal{S})$  can be communicated in a distributed fashion over the memoryless multiple access channel $p(y|x_i ,i\in \mathcal{S})$  with distortions $(D_i ,i\in \mathcal{S})$  if there exist auxiliary random variables $(W_i,X_i ,i\in \mathcal{S})$ satisfying
\begin{eqnarray*}
1.~ p(u_i,z_i,z,w_i,x_i,y ,i \in \mathcal{S}) =   
      p(u_i,z_i,z,i \in \mathcal{S})p(y|x_i,i \in \mathcal{S})\\
 \prod_{j \in \mathcal{S}}p(w_j|u_j,z_j)p(x_j|w_j) 
\end{eqnarray*}
$2.$ There exists a function $f_D: \prod_{j \in \mathcal{S}}\mathcal{W}_j \times  \mathcal{Z} \rightarrow (\hat{\mathcal{U}}_i,i \in \mathcal{S}) $  such that $ E[d(U_{i},\hat{U}_{i})]\leq D_i,~i \in \mathcal{S}$ and the constraints
\begin{equation}
I (U_A,Z_A; W_A | W_{A^c},Z) < I (X_A; Y | X_{A^c}, W_{A^c}, Z) ~\text{for all}~ A \subset \mathcal{S}
\label{eqn1.3}
\end{equation}
are satisfied (in case of continuous channel alphabets we also need the power constraints $E[X_i^2] \leq P_i,~i=1,...,M)$.
\end{theorem}

\subsection{Example}
We provide an example  to show the reduction possible in transmission rates by exploiting the correlation between the sources, the side information and the permissible distortions.

Consider $(U_1, U_2)$ with the joint distribution: 
$P(U_1=0;U_2=0)= P(U_1=1;U_2=1)=1/3;  P(U_1=1;U_2=0)= P(U_1=0;U_2=1)=1/6.$
If we use independent encoders which do not exploit the correlation among the sources then we need $R_1 \geq H(U_1) = 1~bit$ and $R_2 \geq H (U_2) = 1~bit$ for lossless coding of the sources. If we use the coding scheme in \cite{slepian}, then $R_1 \geq H(U_1|U_2) = 0.918~bits, R_2 \geq H (U_2|U_1) = 0.918~bits$ and $R_1+ R_2 \geq H(U_1, U_2) = 1.918~ bits$ suffice. 

Next consider a multiple access channel such that $Y = X_1+ X_2$ where $X_1$ and $X_2$ take values from the alphabet $\{0,1\}$ and $Y$  takes values from the alphabet $\{0,1,2\}$. This does not satisfy the separation conditions (\cite{Medrad06seperation}). The sum capacity $C$ of such a channel with independent $X_1$ and $X_2$ is $1.5~ bits$ and if we use source-channel separation, the given sources cannot be transmitted losslessly because $ H(U_1, U_2) > C$. Now we use a joint source-channel code to improve the capacity of the channel. Take $X_1=U_1$ and $X_2=U_2$. Then the capacity of the channel is improved to $I(X_1,X_2;Y)=1.585~bits$. This is still not enough to transmit the sources over the given MAC. Next we exploit the side information.

The side-information random variables are generated as follows. $Z_1$ is generated from $U_2$ by using a binary symmetric channel (BSC) with cross over probability $p=0.3$. Similarly $Z_2$ is generated from $U_1$ by using the same BSC. Let $Z = (Z_1,Z_2,V)$, where $V= U_1 .U_2 .N$, $N$ is a binary random variable with $P(N=0) = P(N=1) = 0.5 $ independent of $U_1$ and $U_2$ and  `.' denotes the logical AND operation. This  denotes the case when the decoder can observe the encoder side information and also has some extra side information. Then from \eqref{eqn1.1} if we use just the side information $Z_1$ the sum rate for the sources needs to be $1.8~ bits$. By symmetry the same holds if we only have $Z_2$. If we use $Z_1$ and $Z_2$ then we can use the sum rate $1.683~ bits$. If only $V$ is used then the sum rate needed is $1.606~ bits$. So far we can still not transmit $(U_1,U_2)$ losslessly if we use the coding $U_i=X_i,~i=1,2$. If all the information in $Z_1, Z_2,V$ is used then we need $R_1+ R_2 \geq 1.4120 ~bits$. Thus with the aid of $Z_1, Z_2, Z$ we can transmit $(U_1, U_2)$ losslessly over the MAC even with independent $X_1$ and $X_2$.

Next we consider the distortion criterion to be the Hamming distance and the allowable distortion as 4\%. Then for compressing the individual sources without side information we need  $R_i \geq H(p)-H(d) = 0.758~ bits,~i=1,2$, where $H(x)=-xlog_2(x)-(1-x)log_2(1-x)$. Thus we still cannot transmit $(U_1, U_2)$ with this distortion when $(X_1, X_2)$ are independent. If $U_1$ and $U_2$ are encoded, exploiting their correlations, $(X_1, X_2)$ can be correlated.  Next assume the side information $Z=(Z_1,Z_2)$ to be available at the decoder only. Then we need $R_1 \geq  I(U_1;W_1)- I(Z_1; W_1)$ where $W_1$ is an auxiliary random variable generated from $U_1$. $W_1$ and $Z_1$ are related by a cascade of a BSC with crossover probability 0.3 with a BSC with crossover probability 0.04. This implies that $R_1 \geq 0.6577~ bits$ and $R_2 \geq 0.6577~ bits$.
\section{Special Cases}\label{sec1.4}

In the following we provide several systems studied in literature as special cases. The practically important special cases of GMAC and orthogonal channels will be studied in detail in later sections. There we will discuss several specific joint source-channel coding schemes for these and compare their performance.

\subsection{Lossless multiple access communication with correlated sources}
Take   $(Z_1,Z_2,Z) \bot (U_1, U_2)$ ($X \bot Y$ denotes that r.v. $X$ is independent of r.v. $Y$)  and $W_1=U_1$ and $W_2=U_2$ where $U_1,U_2$ are discrete sources. Then the constraints of \eqref{eqn1.1} reduce to 
\begin{eqnarray}
\label{ffirst}
H (U_1| U_2)  &<&  I (X_1; Y | X_2, U_2),\nonumber\\
H(U_2 | U_1)  &<& I (X_2; Y | X_1, U_1),\\	   		    
H(U_1,U_2)  &<&  I (X_1, X_2; Y)\nonumber
\end{eqnarray}
where $X_1$, $X_2$ are the channel inputs, $Y$ is the channel output and $X_1\leftrightarrow U_1 \leftrightarrow U_2\leftrightarrow X_2$ is satisfied. These are the conditions obtained in  \cite{Cover80multiple}. 

\subsection{Lossy multiple access communication}
Take  $(Z_1,Z_2,Z) \bot (U_1, U_2)$ . In this case the constraints in \eqref{eqn1.1} reduce to
\begin{eqnarray}
   I (U_1 ;W_1 | W_2)   &<&  I (X_1;Y | X_2, W_2),\nonumber\\
           I (U_2; W_2| W_1)   &<&  I (X_2;Y |X_1, W_1),\\			  	 
 I (U_1,U_2 ;W_1,W_2)   &<&  I (X_1,X_2;Y).\nonumber
\label{eqna}
\end{eqnarray}
This is an immediate generalization of \cite{Cover80multiple} to the lossy case.

\subsection{Lossy distributed source coding with side information}
The multiple access channel is taken as a dummy channel which reproduces its inputs. In this case we obtain that the sources can be coded with rates $R_1$ and $R_2$ to obtain the specified distortions at the decoder if  
\begin{eqnarray}
R_1  &>&  I (U_1,Z_1; W_1 | W_2,  Z),\nonumber\\
R_2  &>&   I (U_2,Z_2; W_2 | W_1,  Z),\\			     	
R_1  + R_2 &>&  I (U_1, U_2,Z_1,Z_2 ; W_1, W_2 | Z).\nonumber
\label{eqnb}
\end{eqnarray}
This recovers the result in \cite{ncc}, and generalizes the results in \cite{wyner,Gastpar,slepian}.

\subsection{Correlated sources with lossless transmission over multiuser channels with receiver side information}

If we consider $(Z_1,Z_2)\bot (U_1,U_2)$, $W_1=U_1$ and $W_2=U_2$ then we recover the conditions 
\begin{eqnarray}
H(U_1|U_2,Z)  &<&  I (X_1; Y | X_2, U_2, Z),\nonumber\\
H(U_2|U_1,Z)  &<& I (X_2; Y | X_1, U_1, Z),\\	   		    
H (U_1,U_2 |Z) &<& I (X_1, X_2; Y |Z)\nonumber
\label{eqn1.4}
\end{eqnarray}
in $Theorem~2.1$ in \cite{elza}.

\subsection{ Mixed Side Information }
The aim is to determine the rate distortion function for transmitting a source $X$ with the aid of side information $(Y,Z)$ (system in Fig 1(c) of \cite{effros}). The encoder is provided with $Y$ and the decoder has access to both $Y$ and $Z$. This represents the Mixed side information (MSI) system which combines the conditional rate distortion system and the Wyner-Ziv system. This has the system in Fig 1(a) and (b) of \cite{effros} as special cases. The results of Fig 1(c) can be recovered from our Theorem if we take $X,Y,Z,W$ in \cite{effros} as $U_1=X,Z=(Z,Y),Z_1=Y,W_1=W$. $U_2$ and $Z_2$ are taken to be  constants. The acceptable rate region is given by $R > I(X,W|Y,Z)$, where $W$ is a random variable with the property $W\leftrightarrow (X,Y) \leftrightarrow Z$ and for which there exists a decoder function such that the distortion constraints are met.

\section{Discrete Alphabet Sources over Gaussian MAC}\label{sec1.5}
This system is practically very useful. For example, in a sensor network, the observations sensed by the sensor nodes are discretized and then transmitted over a GMAC. The physical proximity of the sensor nodes makes their observations correlated. This correlation can be exploited to compress the transmitted data. We present a distributed `correlation preserving' joint source-channel coding scheme yielding jointly Gaussian channel codewords which will be shown to compress the data efficiently. This coding scheme was developed in \cite{bitg}.

Sufficient conditions for lossless transmission of two discrete sources $(U_1,U_2)$ (generating $iid$ sequences in time) over a general MAC with no side information are obtained in \eqref{ffirst} and reproduced below for convenience
\begin{eqnarray}
\label{first}
H (U_1| U_2)  &<&  I (X_1; Y | X_2, U_2),\nonumber\\
H(U_2 | U_1)  &<& I (X_2; Y | X_1, U_1),\\	   		    
H(U_1,U_2)  &<&  I (X_1, X_2; Y)\nonumber
\end{eqnarray} 
where $X_1\leftrightarrow U_1 \leftrightarrow U_2\leftrightarrow X_2$ is satisfied.

In this section, we further specialize the above results for lossless transmission of discrete correlated sources over an additive memoryless GMAC: $Y = X_1+ X_2+ N $ where $N$ is a Gaussian random variable independent of $X_1$ and $X_2$. The noise $N$ satisfies $E[N] = 0$ and $Var(N)=\sigma_N^2$ . We will also have the transmit power constraints: $E[X_i^2]\leq P_i, i=1,2$. Since source-channel separation does not hold for this system, a joint source-channel coding scheme is needed for optimal performance.

The dependence of R.H.S. of \eqref{first} on input alphabets prevents us from getting a closed form expression for the admissibility criterion. Therefore we relax the conditions by taking away the dependence on the input alphabets. This will allow us to obtain good joint source-channel codes.
\begin{lemma}
 Under our assumptions, $ I(X_1; Y | X_2, U_2) \leq I (X_1; Y | X_2)$.
\label{lemma1}
\end{lemma}
 $Proof$:
  Let 
  \begin{gather} 
   \Delta\buildrel\Delta\over = I ( X_1;Y | X_2, U_2) -  I ( X_1;Y | X_2 ).\nonumber
   \end{gather}
       Then
       \begin{gather}
 \Delta= H(Y | X_2, U_2) - H(Y | X_1, X_2, U_2) \nonumber
 - [H(Y | X_2) - H (Y | X_1, X_2)].\nonumber
 \end{gather}
Since the channel is memoryless,
\begin{gather}
 H(Y | X_1, X_2, U_2) = H(Y | X_1, X_2).\nonumber 
 \end{gather}
 Thus,
     $\Delta = H(Y | X_2, U_2) - H(Y | X_2) \leq 0$. 			            
\begin{flushright}
$\blacksquare$
\end{flushright}
Therefore, from \eqref{first},
\begin{eqnarray}
H (U_1 | U_2)  &<&  I (X_1;Y | X_2, U_2) \leq I ( X_1;Y | X_2 ),\label{sub1}\\
H (U_2 | U_1)  &<& I (X_2;Y | X_1, U_1) \leq  I ( X_2;Y | X_1),\label{sub2}\\
H (U_1, U_2)  &<&   I (X_1, X_2;Y).
\label{eqn0}
\end{eqnarray}
The relaxation of the upper bounds is only in \eqref{sub1} and \eqref{sub2} and not in \eqref{eqn0}.
 
We show that the relaxed upper bounds are maximized if $(X_1, X_2)$ is jointly Gaussian and the correlation $\rho$  between $X_1$ and $X_2$ is high (the highest possible $\rho$  may not give the largest upper bound in the three inequalities in \eqref{sub1}-\eqref{eqn0}). 

\begin{lemma}
A jointly Gaussian distribution for $(X_1, X_2)$ maximizes \\$I (X_1;Y | X_2)$, $I (X_2;Y | X_1)$ and
 $I (X_1, X_2;Y)$ simultaneously.
\label{lemma2}
\end{lemma}
$Proof$: Since
\begin{eqnarray*}
 I (X_1, X_2;Y) &=& H(Y) -  H (Y | X_1, X_2)\\
&=& H (X_1+ X_2+ N) - H (N),
\end{eqnarray*}
it is maximized when $H (X_1+ X_2+ N)$ is maximized. This entropy is maximized when $X_1+ X_2$ is Gaussian with the largest possible variance $=P_1+P_2$. If $(X_1, X_2)$ is jointly Gaussian then so is $X_1+ X_2$.

Next consider $I (X_1;Y | X_2 )$. This equals 
\begin{eqnarray*}
H (Y | X_2) - H (N)&=& H (X_1+ X_2+ N | X_2) - H (N)\\
&=& H (X_1+ N | X_2) - H (N)
\end{eqnarray*}
which is maximized when $P (x_1| x_2)$ is Gaussian and this happens when $X_1, X_2$ are jointly Gaussian.	
			   
A similar result holds for $I (X_2; Y | X_1)$.
\begin{flushright}
$\blacksquare$
\end{flushright}

The difference between the bounds in  \eqref{sub1} is 
\begin{equation}
I(X_1,Y|X_2)-I(X_1,Y|X_2,U_2)=I(X_1+N;U_2|X_2).
\label{on}
\end{equation}
This difference is small if correlation between $(U_1,U_2)$ is small. In that case $H(U_1|U_2)$ and $H(U_2|U_1)$ will be large and \eqref{sub1}  and \eqref{sub2} can be active constraints. If correlation between $(U_1,U_2)$ is large, $H(U_1|U_2)$ and $H(U_2|U_1)$ will be small and \eqref{eqn0} will be the only active constraint. In this case the difference between the two bounds in \eqref{sub1} and \eqref{sub2} is large but not important. Thus, the outer bounds in \eqref{sub1} and \eqref{sub2} are close to the inner bounds whenever the constraints \eqref{sub1} and \eqref{sub2} are active. Often \eqref{eqn0} will be the only active constraint. 

An advantage of outer bounds in \eqref{sub1} and \eqref{sub2} is that we will be able to obtain a good source-channel coding scheme. Once $(X_1,X_2)$ are obtained we can check for sufficient conditions \eqref{first}. If these are not satisfied for the $(X_1,X_2)$ obtained, we will increase the correlation $\rho$ between $(X_1,X_2)$ if possible (see details below). Increasing the correlation in $(X_1,X_2)$ will decrease the difference in \eqref{on} and increase the possibility of satisfying \eqref{first} when the outer bounds in \eqref{sub1} and \eqref{sub2} are satisfied.

We evaluate the (relaxed) rate region \eqref{sub1}-\eqref{eqn0} for the Gaussian MAC with jointly Gaussian channel inputs $(X_1, X_2)$ with the transmit power constraints. For maximization of this region we need mean vector $[0 ~ 0]$ and covariance matrix $K_{X_1,X_2}=\begin{pmatrix}
P_1 & \rho\sqrt{P_1P_2} \\
\rho\sqrt{P_1P_2} &  P_2
\end{pmatrix}$
  where $\rho$ is the correlation between $X_1$ and $X_2$. Then \eqref{sub1}-\eqref{eqn0} provide the relaxed constraints
\begin{eqnarray}
H (U_1 | U_2)  &<&  0.5\log\left[ 1+ \frac{P_1(1- {{\rho}}^{2})} {{\sigma_N}^{2}}\right],\label{coo0}\\
H (U_2 | U_1)  &<& 0.5\log\left[ 1+ \frac{P_2(1- {{\rho}}^{2})} {{\sigma_N}^{2}}\right],\label{coo1}\\   			        
H (U_1, U_2 )  &<&   0.5\log\left[ 1+ \frac{P_1 + P_2 + {2} {{\rho}}{\sqrt{P_1P_2}}} {{\sigma_N}^{2}} \right].  
\label{coo2} 
\end{eqnarray}
 	   	
The upper bounds in the first two inequalities in \eqref{coo0} and \eqref{coo1} decrease as $\rho$  increases. But the third upper bound \eqref{coo2} increases with $\rho$ and often the third constraint is the limiting constraint.

This motivates us to consider the GMAC with correlated jointly Gaussian inputs. The next lemma provides an upper bound on the correlation between $(X_1, X_2)$ in terms of the distribution of $(U_1,U_2)$.

\begin{lemma}
 Let $(U_1, U_2)$ be the correlated sources and $X_1\leftrightarrow U_1\leftrightarrow U_2\leftrightarrow X_2$ where $X_1$ and $X_2$ are jointly Gaussian. Then the correlation   between  $(X_1, X_2)$ satisfies $ \rho^2 \leq 1- 2^{-2I(U_1,U_2)}$.
 \label{lemma3}
 \end{lemma}
$Proof$: Since $X_1\leftrightarrow U_1\leftrightarrow U_2\leftrightarrow X_2$  is a Markov chain, by data processing inequality
$I (X_1; X_2) \leq I (U_1; U_2)$. Taking $X_1, X_2$ to be jointly Gaussian with zero mean, unit variance and correlation $\rho,~I(X_1,X_2)=0.5log_2(\frac{1}{1-\rho^2})$.  This implies $ \rho^2 \leq 1- 2^{-2I(U_1,U_2)}$.
\begin{flushright}
$\blacksquare$
\end{flushright}
\subsection {A coding Scheme}\label{sec1.5.1}

In this section we develop a coding scheme for mapping the discrete alphabets into jointly Gaussian correlated code words which also satisfy the Markov condition. The heart of the scheme is to approximate a jointly Gaussian distribution with the sum of product of Gaussian marginals. Although this is stated in the following lemma for two dimensional vectors $(X_1, X_2)$, the result holds for any finite dimensional vectors.

\begin{lemma}
Any jointly Gaussian two dimensional density can be uniformly arbitrarily closely approximated by a weighted sum of product of marginal Gaussian densities:
\begin{gather}
\sum_{i=1}^N{\frac{p_i}{\sqrt{2\pi c_{1i}}}e^{\frac{-1}{2c_{1i}}(x_1-a_{1i})^2}\frac{q_i}{\sqrt{2\pi c_{2i}}}e^{\frac{-1}{2c_{2i}}(x_2-a_{2i})^2}}.
\label{co2} 
\end{gather}
\label{lemma4}
\end{lemma}
$Proof$: By Stone-Weierstrass theorem (\cite{protter}) the class of functions $(x_1,x_2) \mapsto e^{\frac{-1}{2c_{1}}(x_1-a_{1})^2} e^{\frac{-1}{2c_{2}}(x_2-a_{2})^2}$ can be shown to be dense in $C_0$ under uniform convergence where $C_0$ is the set of all continuous functions on $\Re^2$ such that $\lim_{\|X\| \to \infty}|f(x)|=0$  . Since the jointly Gaussian density \\ $(x_1,x_2) \mapsto e^{\frac{-1}{2\sigma^2}(\frac{x_1^2+x_2^2-2\rho x_1 x_2}{1-\rho^2})}$  is in $C_0$, it can be approximated arbitrarily closely uniformly by the functions \eqref{co2}.
\begin{flushright}
$\blacksquare$
\end{flushright}

	From the above lemma we can form a sequence of functions $f_n(x_1,x_2)$ of type \eqref{co2} such that $sup_{x_1,x_2}|f_n(x_1,x_2)-f(x_1,x_2)| \rightarrow 0$ as $ n \rightarrow \infty$, where $f$ is a given jointly Gaussian density. Although $f_n$ are not guaranteed to be probability densities, due to uniform convergence, for large $n$, they will almost be. In the following lemma we will assume that we have made the minor modification to ensure that $f_n$ is a proper density for large enough $n$. This lemma shows that obtaining $(X_1, X_2)$ from such approximations can provide the (relaxed) upper bounds in (2)-(4) (we actually show for the third inequality only but this can be shown for the other inequalities in the same way). Let $(X_{m1},X_{m2})$  and $(X_1, X_2)$ be random variables with distributions $f_m$ and $f$ and $sup_{x_1,x_2}|f_m(x_1,x_2)-f(x_1,x_2)|\rightarrow0$ as $ m \rightarrow \infty$. Let $Y_m$ and $Y$ denote the corresponding channel outputs.

\begin{lemma}
For the random variables defined above, if $\{logf_m(Y_m),m\geq1\}$  is uniformly integrable, $I(X_{m1},X_{m2};Y_m) \rightarrow I(X_1,X_2;Y)$ as $m \rightarrow \infty$ .
\end{lemma} 
$Proof$: Since
\begin{eqnarray*} 
I(X_{m1},X_{m2};Y_m)&=& H(Y_m)-H(Y_m|X_{m1},X_{m2}\\
&=& H(Y_m)-H(N),
\end{eqnarray*}
it is sufficient to show that $H(Y_m) \rightarrow H(Y)$.

From $(X_{m1},X_{m2}) {\buildrel d \over \longrightarrow} (X_1,X_2)$  and independence of $(X_{m1},X_{m2})$   from $N$, we get 
 $ Y_m= X_{m1}+ X_{m2} + N {\buildrel d \over \longrightarrow} X_1+X_2+N = Y$. Then $f_m \rightarrow f$  uniformly implies that $f_m(Y_m){\buildrel d \over \longrightarrow} f(Y)$. Since $f_m(Y_m) \geq 0,~ f(Y) \geq 0~ a.s $  and $log$ is continuous except at $0$, we  obtain $ logf_m(Y_m){\buildrel d \over \longrightarrow} logf(Y)$  . Then uniform integrability provides $I(X_{m1},X_{m2};Y_m) \rightarrow I(X_1,X_2;Y)$.
\begin{flushright}
$\blacksquare$
\end{flushright}

A set of sufficient conditions for uniform integrability of $\{logf_m(Y_m),m\geq1\}$  is 		
\begin{enumerate}
	\item  Number of components in \eqref{co2} is upper bounded.
	\item Variance of component densities in \eqref{co2} is upper bounded and lower bounded away from zero.
	\item The means of the component densities in \eqref{co2} are in a bounded set.
\end{enumerate}
 						 
From Lemma \ref{lemma4} a joint Gaussian density with any correlation  can be expressed by a linear combination of marginal Gaussian densities. But the coefficients $p_i$ and $q_i$ in \eqref{co2} may be positive or negative. To realize our coding scheme, we would like to have the $p_i$'s and $q_i$'s to be non negative. This introduces constraints on the realizable Gaussian densities in our coding scheme. For example, from Lemma \ref{lemma3}, the correlation $\rho$  between $X_1$ and $X_2$ cannot exceed $\sqrt{1-2^{-2I(U_1;U_2)}}$. Also there is still the question of getting a good linear combination of marginal densities to obtain the joint density for a given $N$ in \eqref{co2}.

This motivates us to consider an optimization procedure for finding $p_i,q_i,a_{1i},a_{2i},c_{1i}$ and $c_{2i}$ in \eqref{co2} that provides the best approximation to a given joint Gaussian density. We illustrate this with an example. Consider $U_1, U_2$ to be binary. Let $ P(U_1=0; U_2=0)=p_{00}; P(U_1=0; U_2=1)=p_{01};  P(U_1=1; U_2=0)=p_{10} $ and $ P(U_1=1; U_2=1)= p_{11}$. We can consider
\begin{gather}
f(X_1=.|U_1=0)=p_{101}\mathcal{N}(a_{101},c_{101})+p_{102}\mathcal{N}(a_{102},c_{102})\nonumber\\
...+p_{10r_1}\mathcal{N}(a_{10r_1},c_{10r_1}),\\ 
f(X_1=.|U_1=1)=p_{111}\mathcal{N}(a_{111},c_{111})+p_{112}\mathcal{N}(a_{112},c_{112})\nonumber\\
...+p_{11r_2}\mathcal{N}(a_{11r_2},c_{11r_2}),\\
f(X_2=.|U_2=0)=p_{201}\mathcal{N}(a_{201},c_{201})+p_{202}\mathcal{N}(a_{202},c_{202})\nonumber\\
...+p_{20r_3}\mathcal{N}(a_{20r_3},c_{20r_3}),\\
f(X_2=.|U_2=1)=p_{211}\mathcal{N}(a_{211},c_{211})+p_{212}\mathcal{N}(a_{212},c_{212})\nonumber\\
...+p_{21r_4}\mathcal{N}(a_{21r_4},c_{21r_4}).
\label{cond}
\end{gather} 
where $\mathcal{N}(a,b)$  denotes Gaussian density with mean $a$  and variance $b$. Let $\underline{p}$  be the vector with components $p_{101},...,p_{10r_1}$,$p_{111},...,p_{11r_2}$,$p_{201},...,p_{20r_3}$, $p_{211}$, $...$, $p_{21r_4}$. Similarly we denote by $\underline{a}$  and $\underline{c}$ the vectors with components $a_{101},...,a_{10r_1}$, $a_{111},...,a_{11r_2}$, $a_{201},...,a_{20r_3}$, $a_{211},...,a_{21r_4}$ and $c_{101},...,c_{10r_1}$, $c_{111},...,c_{11r_2}$, $c_{201},...,c_{20r_3}$, $c_{211},...,c_{21r_4}$.

Let $f_\rho(x_1,x_2)$ be the jointly Gaussian density that we want to approximate. Let it has zero mean and covariance matrix $K_{X_1,X_2}=\begin{pmatrix}
1 & \rho \\
\rho &  1
\end{pmatrix}$. Let $g_{\underline{p},\underline{a},\underline{c}}$ be the sum of marginal densities with parameters $\underline{p},\underline{a},\underline{c}$ approximating $f_\rho$ . The best $g$ is obtained by solving the following minimization problem:
\begin{gather}
\text{min}_{\underline{p},\underline{a},\underline{c}}\int{[g_{\underline{p},\underline{a},\underline{c}}(x_1,x_2)- f_\rho(x_1,x_2)]^2 dx_1dx_2}
\label{optim}
\end{gather}
subject to 
\begin{gather}
(p_{00}+p_{01})\sum_{i=1}^{r_1}p_{10i}a_{10i}+(p_{10}+p_{11})\sum_{i=1}^{r_2}p_{11i}a_{11i}=0,\nonumber\\
(p_{00}+p_{10})\sum_{i=1}^{r_3}p_{20i}a_{20i}+(p_{01}+p_{11})\sum_{i=1}^{r_4}p_{21i}a_{21i}=0,\nonumber
\end{gather}
\begin{gather}
(p_{00}+p_{01})\sum_{i=1}^{r_1}p_{10i}(c_{10i}+a_{10i}^2)+\nonumber (p_{10}+p_{11})\sum_{i=1}^{r_2}p_{11i}(c_{11i}+a_{11i}^2)=1,\nonumber
\end{gather}
\begin{gather}
(p_{00}+p_{10})\sum_{i=1}^{r_3}p_{20i}(c_{20i}+a_{20i}^2)+\nonumber
(p_{01}+p_{11})\sum_{i=1}^{r_4}p_{21i}(c_{21i}+a_{21i}^2)=1,\nonumber
\end{gather}
\begin{gather}
\sum_{i=1}^{r_1}p_{10i}=1,\sum_{i=1}^{r_2}p_{11i}=1,\sum_{i=1}^{r_3}p_{20i}=1,\sum_{i=1}^{r_4}p_{21i}=1,\nonumber
\end{gather}
\begin{gather}
p_{10i}\geq 0, c_{10i}\geq 0 ~for~i \in \{1,2...r_1\},\nonumber 
p_{11i}\geq 0, c_{11i}\geq 0 ~for~ i \in \{1,2...r_2\},\nonumber \\
p_{20i}\geq 0, c_{20i}\geq 0~ for~i \in \{1,2...r_3\},\nonumber 
p_{21i}\geq 0, c_{21i}\geq 0 ~for~ i \in \{1,2...r_4\}.\nonumber 
\label{const}
\end{gather}
The above constraints are such that the resulting distribution $g$ for $(X_1,X_2)$ will satisfy $E[X_i]=0$ and $E[X_i^2]=1,~i=1,2$.

The above coding scheme will be used to obtain a codebook as follows. If user 1 produces $U_1=0$, then with probability $p_{10i}$
  the encoder 1 obtains codeword $X_1$   from the distribution $\mathcal{N}(a_{10i},c_{10i})$. Similarly we obtain the codewords for $U_1=1$ and for user 2. Once we have found the encoder maps the encoding and decoding are as described in the proof of Theorem 1 in Appendix A. The decoding is done  by joint typicality of the received $Y^n$ with $(U_1^n,U_2^n)$. 

This coding scheme can be extended to any discrete alphabet case. We give an example below to illustrate the coding scheme.

\subsection{Example }
Consider $(U_1, U_2)$ with the joint distribution: 
$P(U_1=0; U_2=0) = P(U_1=1; U_2=1)= P(U_1=0; U_2=1)=1/3;P(U_1=1; U_2=0)=0$
and power constraints $P_1 = 3 ; P_2 = 4$. Also consider a Gaussian multiple access channel with $\sigma_N^2 =1$. If the sources are mapped into independent channel code words, then the sum rate condition in \eqref{coo2} with $\rho=0$  should hold. The LHS evaluates to 1.585 bits whereas the RHS is 1.5 bits. Thus condition \eqref{coo2} is violated and hence the sufficient conditions in \eqref{first} are also violated. 

\begin{figure}[h]
\centering
\includegraphics [width=2.4in,height=2in] {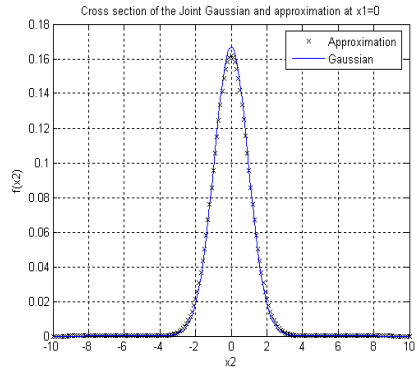}\label{figapprox1}
\caption{Cross section of the approximation of the joint Gaussian $\rho$=0.3}
\end{figure}

In the following we explore the possibility of using correlated $(X_1, X_2)$ to see if we can transmit this source on the given MAC. The inputs $(U_1, U_2)$ can be distributedly mapped to jointly Gaussian channel code words $(X_1, X_2)$ by the technique mentioned above. The maximum $\rho$ which satisfies \eqref{coo0} and \eqref{coo1} are 0.7024 and 0.7874 respectively and the minimum $\rho$  which satisfies \eqref{coo2} is 0.144. Thus, we can pick a $\rho$  which satisfies \eqref{coo0}-\eqref{coo2}. From Lemma~\ref{lemma3}, $\rho$ is upper bounded by 0.546. Therefore we want to obtain jointly Gaussian $(X_1, X_2)$ satisfying $X_1\leftrightarrow U_1\leftrightarrow U_2\leftrightarrow X_2$ with correlation $\rho \in [0.144,0.546]$. If we pick a $\rho$ that satisfies the original bounds, then  we will be able to transmit the sources $(U_1, U_2)$ reliably on this MAC. Without loss of generality the jointly Gaussian channel inputs required are chosen with mean vector $[0~ 0]$ and covariance matrix $K_{X_1,X_2}=\begin{pmatrix}
1 & \rho \\
\rho &  1
\end{pmatrix}$. The $\rho$  chosen is 0.3 and hence is such that it meets all the conditions \eqref{coo0}-\eqref{coo2}. Also, we choose $r_1 = r_2 = r_3 = r_4= 2$. We solve the  optimization problem \eqref{optim} via MATLAB to get the function $g$. The normalized minimum distortion, defined as ${\int{[g_{\underline{p},\underline{a},\underline{c}}(x_1,x_2)- f_\rho(x_1,x_2)]^2 dx_1dx_2}}/{\int{f_\rho^2(x_1,x_2)dx_1dx_2}}$ is 0.137\% when the marginals are chosen as:
\begin{gather}
f(X_1|U_1=0)=\mathcal{N}(-0002,0.9108),~\nonumber
f(X_1|U_1=1)=\mathcal{N}(-0001,1.0446),\nonumber\\
f(X_2|U_2=0)=\mathcal{N}(-0021,1.1358),~\nonumber
f(X_2|U_2=1)=\mathcal{N}(-0042,0.7283).\nonumber
\label{resul}
\end{gather}
The approximation (a cross section of the two dimensional densities) is shown in Fig.~\ref{figapprox1}.

\begin{figure}[h]
\centering
\includegraphics [width=2.4in,height=2in] {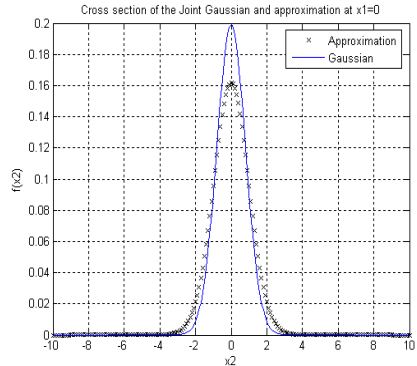}\label{figapprox2}
\caption{Cross section of the approximation of the joint Gaussian $\rho$=0.6}
\end{figure}

If we take $\rho =0.6$ which violates Lemma~\ref{lemma3} then the approximation is shown in Fig.~\ref{figapprox2}.  We can see from Fig.~\ref{figapprox2} that the error in this case is more. Now the normalized marginal distortion is 10.5 \%.

 The original upper bound in \eqref{sub1} and \eqref{sub2} for this example with $\rho=0.3$ is $I (X_1;Y | X_2, U_2) =0.792,~I (X_2;Y | X_1, U_1)=0.996$. Also, $I ( X_1;Y | X_2 )=0.949,~ I ( X_2;Y | X_1)=1.107$. $H(U_1|U_2)=H(U_2|U_1)=0.66$ and we conclude that the original bounds too are satisfied by the choice of $\rho=0.3$. 

\section{Source-Channel Coding for Gaussian sources over Gaussian MAC}\label{sec1.6}
In this section we consider transmission of correlated Gaussian sources over a GMAC. This is an important example for transmitting continuous alphabet sources over a GMAC. For example one comes across it if a sensor network is sampling a Gaussian random field. Also, in the application of detection of change (\cite{Veeravalli01decentralised}) by a sensor network, it is often the detection of change in the mean of the sensor observations with the sensor observation noise being Gaussian.

We will  assume that $(U_{1n}, U_{2n})$ is jointly Gaussian with mean zero, variances $\sigma_i^2,~i=1,2$   and correlation $ \rho. $ The distortion measure will be Mean Square Error (MSE). The (relaxed) sufficient conditions from \eqref{eqna} for transmission of the sources over the channel are given by (these continue to hold because Lemmas~\ref{lemma1}-\ref{lemma3} are still valid)
\begin{eqnarray}
I (U_1; W_1 | W_2)  &<&  0.5\log\left[ 1+ \frac{P_1(1- {\tilde{\rho}}^{2})} {{\sigma_N}^{2}}\right]\nonumber,\\
I (U_2; W_2 | W_1)   &<&   0.5\log\left[ 1+ \frac{P_2(1- {\tilde{\rho}}^{2})} {{\sigma_N}^{2}}\right],\label{gogmac}\\   			        
I (U_1, U_2 ; W_1, W_2)   &<&    0.5\log\left[ 1+ \frac{P_1 + P_2 + {2} {\tilde{\rho}}{\sqrt{P_1P_2}}} {{\sigma_N}^{2}} \right]\nonumber.   
\end{eqnarray}
We consider three specific coding schemes to obtain $W_1,W_2,X_1,X_2$  where $(W_1,W_2)$   satisfy the distortion constraints and $(X_1,X_2)$ are jointly Gaussian with an appropriate $\tilde{\rho}$  such that \eqref{gogmac} is satisfied. These coding schemes have been widely used. We compare their performance also.

\subsection {Amplify and forward scheme}\label{sec1.6.1}

In the Amplify and Forward (AF) scheme the channel codes $X_i$ are just scaled source symbols $U_i$. Since $(U_1,U_2)$ are themselves jointly Gaussian, $(X_1,X_2)$ will be jointly Gaussian and retain the dependence of inputs $(U_1,U_2).$ The scaling is done to ensure $E[{X_i}^{2}]= P_i,i=1,2$. For a single user case this coding is optimal \cite{Cover04elements}.

 At the decoder inputs $U_1$ and $U_2$ are directly estimated from $Y$ as $\hat{U_i}=E[U_i|Y],~i=1,2$. Because $U_i$ and $Y$ are jointly Gaussian this estimate is linear and also satisfies the Minimum Mean Square Error (MMSE) and Maximum Likelihood (ML) criteria. 

The MMSE distortion for this encoding-decoding scheme is
\begin{equation}
 \overline{D_1} = \frac{{\sigma_1}^{2} \left [P_2(1-\rho^{2})+ {\sigma_N}^{2}\right]} {P_1 + P_2 + {2} {\rho}{\sqrt{P_1P_2}}+{\sigma_N}^{2} },
  \overline{D_2}= \frac{{\sigma_2}^{2} \left [P_1(1-\rho^{2})+ {\sigma_N}^{2}\right]} {P_1 + P_2 + {2} {\rho}{\sqrt{P_1P_2}}+{\sigma_N}^{2} }.
\label{dist}
\end{equation}
  
Since encoding and decoding require minimum processing and delay in this scheme, if it satisfies the required distortion bounds $D_i$, it should be the scheme to implement. This scheme has been studied in  \cite{Lapidoth01sending} and found to be optimal below a certain SNR for two-user symmetric case $(P_1=P_2,\sigma_1=\sigma_2,D_1=D_2).$ However unlike for single user case, in this case user 1 acts as interference for user 2 (and vice versa). Thus one should not expect this scheme to be optimal under high SNR case. That this is indeed true was shown in Ref.\cite{Rajesh07allerton}. It was also shown there, that at high SNR, for $P_1 \ne P_2$, it may indeed be better in AF to use less power than $P_1,P_2$. This can also be interpreted as using AF on $U_1-\alpha_1E[U_2|U_1]$  and $U_2-\alpha_2E[U_1|U_2]$ at the two encoders at high SNR which will reduce the correlations between the transmitted symbols.

\subsection{Separation based scheme}\label{sec1.6.2}
\begin{figure*}[ht]
\centering
\includegraphics [height=1.25in, width=1\textwidth] {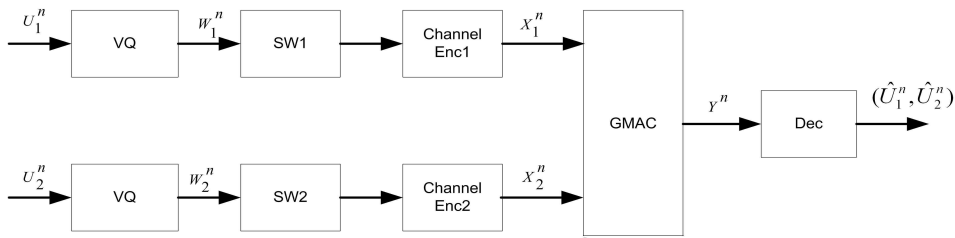}
\caption{separation based scheme}
\label{sb}
\end{figure*}
In separation based (SB) approach (Fig.~\ref{sb}) the jointly Gaussian sources are vector quantized to $W_1^n$ and $W_2^n$. The quantized outputs are Slepian-Wolf encoded \cite{slepian}. This produces code words, which are (asymptotically) independent. These independent code words are encoded to capacity achieving Gaussian channel codes $(X_1^n,X_2^n)$ with correlation  $\tilde{\rho} = 0$. This is a very natural scheme and has been considered by various authors (\cite{Cover80multiple},\cite{Medrad06seperation},\cite{Cover04elements}).

	Since source-channel separation does not hold for this system, this scheme is not expected to be optimal. But because  this scheme decouples source coding from channel coding, it is preferable to a joint source-channel coding scheme with comparable performance.
\subsection{Lapidoth-Tinguely scheme}\label{sec1.6.3}

In this scheme, obtained in \cite{Lapidoth01sending}, $(U_1^n,U_2^n)$ are vector quantized to $2^{nR_1},~2^{nR_2}$ $(\tilde{U}_1^n,\tilde{U}_2^n)$ vectors where $R_1$ and $R_2$ will be specified below. Also, $W_1^n, W_2^n$ are $2^{nR_1}$ and $2^{nR_2}$, $n$ length code words obtained independently with distributions $\mathcal{N}(0,1)$. For each $\tilde{u}_i^n$, we pick the codeword $w_i^n$ that is closest to it. This way we obtain Gaussian codewords $W_1^n, W_2^n$ which retain the correlations of $(U_1^n,U_2^n)$.  $X_1^n$ and $X_2^n$ are obtained by scaling $W_1^n, W_2^n$ to satisfy the transmit power constraints. We will call this LT scheme.  $(U_1,U_2,W_1,W_2)$  are (approximately) jointly Gaussian with covariance matrix 
 \begin{figure*}[ht]
\centering
\begin{equation}
\begin{pmatrix}
\sigma_1^2& \rho\sigma_1\sigma_2& \sigma_1^2(1-2^{-2R_1})& \rho\sigma_1\sigma_2(1-2^{-2R_2}) \\
\rho\sigma_1\sigma_2& \sigma_2^2& \rho\sigma_1\sigma_2(1-2^{-2R_1})& \sigma_2^2(1-2^{-2R_2}) \\
\sigma_1^2(1-2^{-2R_1})& \rho\sigma_1\sigma_2(1-2^{-2R_1})& \sigma_1^2(1-2^{-2R_1})& \frac{\tilde{\rho}^2\sigma_1\sigma_2}{\rho}\\
\rho\sigma_1\sigma_2(1-2^{-2R_2})& \sigma_2^2(1-2^{-2R_2})& \frac{\tilde{\rho}^2\sigma_1\sigma_2}{\rho}& \sigma_2^2(1-2^{-2R_2})
\end{pmatrix}.
\label{covmat}
\end{equation}
\end{figure*}

In \eqref{covmat} $\widetilde{\rho}=\rho\sqrt{(1-2^{-2R_1})(1-2^{-2R_2})}$.

We obtain the $(R_1,R_2)$ above from \eqref{gogmac}. From
\begin{equation*}
 I(U_1;W_1|W_2)  = H (W_1| W_2) - H(W_1| W_2, U_1),
 \end{equation*}
and the fact that the  Markov chain  condition $W_1\leftrightarrow U_1\leftrightarrow U_2\leftrightarrow W_2$   holds,
\begin{equation*}
H(W_1 | W_2, U_1) = H(W_1 | U_1)
\end{equation*}
 and
\begin{equation*}
I (U_1;W_1 | W_2)  = 0.5\log\left[ (1- {\tilde{\rho}}^{2})2^{2R_1}\right].
\end{equation*}
Thus from  \eqref{gogmac} we need $R_1$ and $R_2$ which satisfy
\begin{equation}
 R_1   \leq  0.5\log\left[ \frac{P_1}{{\sigma_N}^{2}}+ \frac{1}{(1- {\tilde{\rho}}^{2})}\right].
\label{const3}
\end{equation}
Similarly, we also need 

\begin{eqnarray}
R_2   &\leq&  0.5\log\left[ \frac{P_2}{{\sigma_N}^{2}}+ \frac{1}{(1- {\tilde{\rho}}^{2})}\right] ,\label{cont4}\\
R_1+R_2  &\leq&    0.5\log\left[ \frac{{\sigma_N}^{2}+ P_1 + P_2 + {2} {\tilde{\rho}}{\sqrt{P_1P_2}}} {{(1- {\tilde{\rho}}^{2})}{\sigma_N}^{2}} \right].   
\label{const4}
\end{eqnarray}

The inequalities \eqref{const3}-\eqref{const4} are the same as in \cite{Lapidoth01sending}. Thus we recover the conditions in \cite{Lapidoth01sending} from our general result (\eqref{eqn1.1}). Taking $\hat{U}_i=E[U_i|W_1,W_2],~i=1,2$, we obtain the distortions
\begin{eqnarray}
D_1 =  var(U_1|W_1,W_2)&=& 
\frac{{\sigma_1}^{2}2^{-2R_1}\left[1-\rho^{2}\left(1-2^{-2R_2}\right)\right]}{(1- {\tilde{\rho}}^{2})},\label{const5}\\
D_2 = var(U_2|W_1,W_2)&=& 
\frac{{\sigma_2}^{2}2^{-2R_2}\left[1-\rho^{2}\left(1-2^{-2R_1}\right)\right]}{(1- {\tilde{\rho}}^{2})}.
\label{const6}
\end{eqnarray}

The minimum distortion is obtained when $\tilde{\rho}$ is such that the sum rate is met with equality in \eqref{const4}. For the symmetric case at the minimum distortion, $R_1=R_2$.

\subsection{Asymptotic performance of the three schemes} 
We compare the performance of the three schemes. These results are from \cite{Rajesh07allerton}. For simplicity we consider the symmetric case: $P_1=P_2=P$, $ \sigma_1=\sigma_2=\sigma $, $D_1=D_2=D$. We will denote the SNR  ${P}/{{\sigma_N}^{2}}$  by $S$.
	
Consider the AF scheme. From \eqref{dist} 
	
\begin{equation}
D(S)=\frac{\sigma^{2}\left[S\left(1-\rho^2\right)+1\right]}{2S\left(1+\rho\right)+1}.
\label{asym1}
\end{equation}
Thus $D(S)$  decreases to ${\sigma^2(1-\rho)}/{2}$   strictly monotonically at rate $O(1)$ as $ S \rightarrow \infty$.

Also,
\begin{equation}
	\lim_{S \to 0}\left|{\frac{D(S)-\sigma^2}{S}}\right|= \sigma^2(1+\rho)^2.
\label{NEW}
 \end{equation}
Thus, $ D(S)\rightarrow \sigma^2$  at rate $O(S)$   as $S \rightarrow 0$ .\\

Consider the SB scheme at High SNR. From \cite{Wagner05rate} if each source is encoded with rate $R$ then it can be decoded at the decoder with distortion	
\begin {equation}
D^2 = 2^{-4R}(1-\rho^2)+\rho^2 2^{-8R}.	
\label{asym2}
\end{equation}		
At high SNR, from the capacity result for independent inputs, we have  $ R < 0.25 \log  S $ (\cite{Cover04elements}). Then from \eqref{asym2} we obtain
 
\begin {equation}
D \geq \sqrt{\frac{\sigma^4(1-\rho^2)}{S}+ \frac{\sigma^4\rho^2}{S^2}}
\label{asym3}	
\end{equation}  	      		  	                 
and this lower bound is achievable. As $S \rightarrow \infty$, this lower bound approaches zero at rate $O(\sqrt{S})$. Thus  this scheme outperforms AF at high SNR.

At low SNR, $R\approx \frac{S}{2}$ and hence from \eqref{asym2}

\begin{equation}
D\geq \rho^2\sigma^42^{-4S} + \sigma^2(1-\rho^2)2^{-2S}.
\label{new2}
\end{equation}
Thus   $D \rightarrow \sigma^2$  at rate $O(S^2)$  as  $S \rightarrow 0 $   at high $\rho$ and at rate $O(S)$   at small $\rho$. Therefore we expect that at low SNR, at high $\rho$ this scheme will be worse than AF but at low $\rho$  it will be comparable.

Consider the LT scheme. In the high SNR region we assume that $\tilde{\rho}= \rho$   since $R = R_1 = R_2$ are sufficiently large. Then from \eqref{const4} $R \approx 0.25log[ 2S/(1-\rho)]$ and the distortion can be approximated by 
\begin{equation}
 D\approx \sigma^2 \sqrt{(1-\rho)/2S}.
 \label{asym4}
 \end{equation}
 Therefore, $D \rightarrow 0 $ as $S \rightarrow \infty $ at rate $O(\sqrt{S} )$. This rate of convergence is same as for SB. However, the R.H.S. in \eqref{asym3} is greater than that of  \eqref{asym4} and at low  $\rho$ the two are close. Thus at high SNR LT always outperforms SB but the  improvement is small for low $\rho$.\\

At low SNR
\begin{equation*}
	R \approx \frac {S(1+ \tilde{\rho})}{2} - \frac{\log(1-\tilde\rho^2)}{4} 
\end{equation*}
and evaluating $D$ from \eqref{const5} we get
	
\begin {equation}
D= \frac{ \sigma^2 2^{-\overline{S}}\left(1-\rho^2(1-\sqrt{1-\tilde{\rho}^2}2^{-\overline{S}})\right)}{\sqrt{1-\tilde{\rho}^2}}	
\label{asym5}
\end{equation} 
where $\overline{S}= S(1+\tilde{\rho})$. Therefore  $D \rightarrow \sigma^2$  as $S \rightarrow 0 $  at rate $O(S^2)$  at high  $\rho$ and at rate $O(S)$ at low $\rho$. These rates are same as that for SB. In fact, dividing the expression for $ D$ at low SNR for SB by that for LT, we can show that the two distortions tend to  $\sigma^2$ at the same rate for all $\rho$ .

The necessary conditions (NC) to be able to transmit on the GMAC with distortion $(D, D)$ for the symmetric case are (\cite{Lapidoth01sending})

\begin{eqnarray}
D\geq \begin{cases}
\frac{\sigma^{2}\left[S\left(1-\rho^2\right)+1\right]}{2S\left(1+\rho\right)+1}~,& \text{for $S \leq \frac{\rho}{1-\rho^2}$},\\\\
\sigma^2\sqrt{\frac{(1-\rho^2)}{2S(1+\rho)+1}}~,& \text{for $S >\frac{\rho}{1-\rho^2}$}.
\end{cases}
\label{asym6}
\end{eqnarray}

\begin{figure}[h]
\centering
\includegraphics [width=2.4in,height=2in] {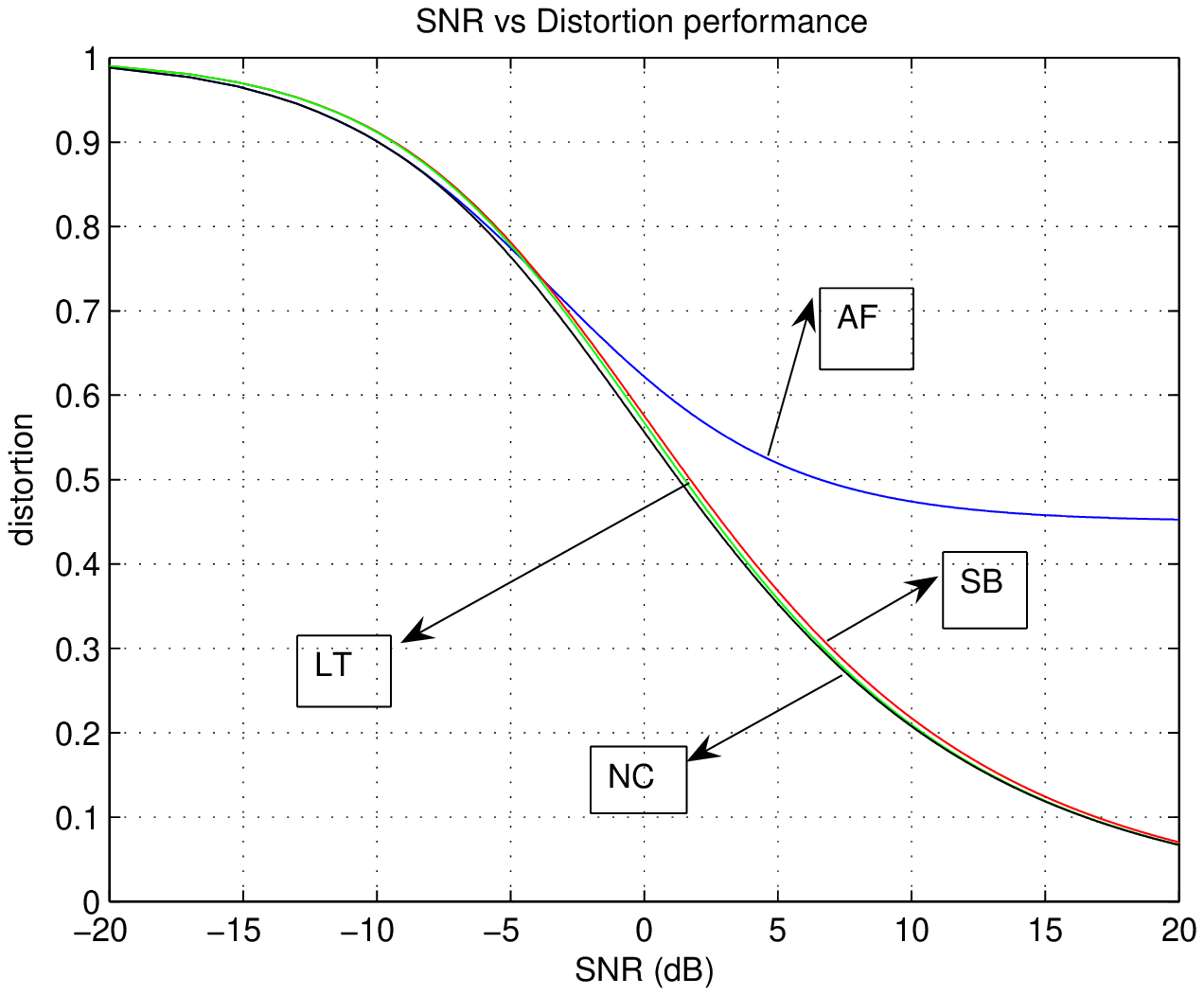}
\caption{SNR vs distortion performance $\rho$=0.1}
\label{fig1.4}
\end{figure}

\begin{figure}[h]
\centering
\includegraphics [width=2.4in,height=2in] {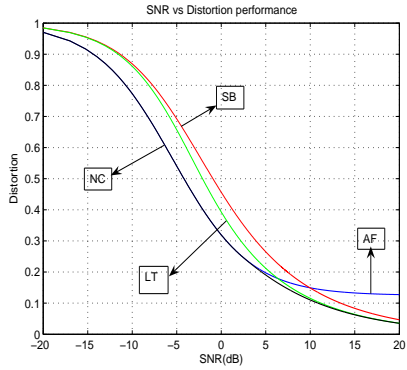}
\caption{SNR vs distortion performance $\rho$=0.75}
\label{fig1.5}
\end{figure}

The above three schemes along with \eqref{asym6} are compared below using exact computations. Figures \ref{fig1.4} and \ref{fig1.5}  shows the distortion as a function of SNR for unit variance jointly Gaussian sources with correlations $\rho$  = 0.1 and 0.75. 

From these plots we confirm our theoretical conclusions provided above. 

\subsection{Continuous sources over a GMAC}
For general continuous alphabet sources $(U_1,U_2)$ we vector quantize $U_1^n,U_2^n$ into $\tilde{U}_1^n,\tilde{U}_2^n$. Then to obtain correlated Gaussian codewords $(X_1^n,X_2^n)$ we can use two schemes adapted from the cases studied above. In the first scheme we use the scheme developed in sec.~\ref{sec1.5.1}. In the second scheme we use LT scheme explained in sec.~\ref{sec1.6.3}.

\section{Correlated Sources over Orthogonal MAC}\label{sec1.7}
One standard way to use the MAC is via TDMA, FDMA, CDMA or Orthogonal Frequency Division Multiple Access (OFDMA) (\cite{Cover04elements,pbook,barros}). These protocols although suboptimal are used due to practical considerations. These protocols make a MAC a set of parallel orthogonal channels (for CDMA, it happens if we use orthogonal codes). We study transmission of correlated sources through such a system.

\subsection {Transmission of correlated sources over orthogonal  channels}\label{sec1.7.1}

Consider the setup in Fig.~\ref{fig1.1} when $Y=(Y_1,Y_2)$ and $p(y|x_1,x_2) $ = $p(y_1,y_2|x_1,x_2)$ = $ \\ p(y_1|x_1)p(y_2|x_2)$. Then the conditions in \eqref{eqn1.1} become

\begin{eqnarray}
\label{conints2}  
I (U_1,Z_1; W_1 | W_2,Z)  &<&  I (X_1; Y_1 | W_2,Z) \leq I(X_1;Y_1),\label{ap1}\\
I (U_2,Z_2;W_2 |W_1,Z)  &<& I (X_2; Y_2 | W_1,Z)\leq I(X_2;Y_2),\label{ap2}\\ 	
I (U_1, U_2,Z_1,Z_2 ; W_1, W_2|Z)  &<&   I (X_1, X_2; Y_1,Y_2|Z)\leq I(X_1;Y_1)+I(X_2;Y_2).\nonumber\\
\label{ap3} 
\end{eqnarray}
The outer bounds in \eqref{ap1}-\eqref{ap3} are attained if the channel codewords $(X_1,X_2)$ are independent of each other. Also, the distribution of $(X_1,X_2)$ maximizing these bounds are not dependent on the distribution of $(U_1,U_2)$. This implies that source-channel separation holds for this system even with side information $(Z_1,Z_2,Z)$ (for the sufficient conditions \eqref{eqn1.1}). Thus by choosing $(X_1,X_2)$ which maximize the outer bounds in \eqref{ap1}-\eqref{ap3} we obtain capacity region for this system which is independent of the side conditions. Also, for a GMAC this is obtained by independent Gaussian r.v.s $X_1$ and $X_2$ with distributions $\mathcal{N}(0,P_i),~i=1,2$, where $P_i$ are the power constraints. Furthermore, the L.H.S. of the inequalities are simultaneously minimized when $W_1$ and $W_2$ are independent. Thus, the source coding $(W_1,W_2)$ on $(U_1,Z_1)$ and $(U_2,Z_2)$ can be done as in Slepian-Wolf coding (by first vector quantizing in case of continuous valued $U_1,U_2$) but also taking into account the fact that the side information $Z$ is available at the decoder. In this section this coding scheme will be called SB.

If we take $W_1=U_1$ and $W_2=U_2$ and  the side information $(Z_1,Z_2,Z)$  $\bot$ $(U_1,U_2)$, we can recover the conditions in \cite{barros}.
 
\subsection {Gaussian sources and orthogonal Gaussian channels}

Now we consider the transmission of jointly Gaussian sources over orthogonal Gaussian channels. Initially it will also  be assumed that there is no side information $(Z_1,Z_2,Z)$.

Now $(U_1,U_2)$ are zero mean jointly Gaussian random variables with variances $\sigma_1^2$ and $\sigma_2^2$ respectively and correlation $\rho$. Then $Y_i=X_i+N_i, i=1,2$ where $N_i$ is Gaussian with zero mean and $\sigma_{N_i}^2$ variance. Also $N_1$ and $N_2$ are independent of each other and also of $(U_1,U_2)$. 

In this scenario, the R.H.S. of the inequalities in  \eqref{ap1}-\eqref{ap3} are maximized by taking $X_i \sim \mathcal{N}(0,P_i),~i=1,2$ independent of each other where $P_i$ is the average transmit power constraint on user $i$. Then $I(X_i,Y_i)=0.5log(1+P_i/\sigma_{N_i}^2),~i=1,2$.

Based on the comments at the end of sec.~\ref{sec1.7.1}, for two users, using the results from \cite{Wagner05rate} we obtain the necessary and sufficient conditions for transmission on an orthogonal GMAC with given distortions $D_1$ and $D_2$.

We can specialize the above results to a TDMA, FDMA or CDMA  based transmission scheme. The specialization to TDMA is given here. Suppose source 1  uses the channel $\alpha$ fraction of time and user 2, $1-\alpha$ fraction of time. In this case we can use average power $P_1/ \alpha$ for the first user and  $ P_2/(1-\alpha)$ for the second user whenever they transmit. The conditions  \eqref{ap1}-\eqref{ap3} for the optimal scheme become

\begin{equation}
I (U_1; W_1 | W_2)  < 0.5\alpha\log\left[ 1+ \frac{P_1} {{\alpha\sigma_{N_1}}^{2}}\right],
\label{or1}
\end{equation}
\begin{equation}         
I (U_2; W_2 | W_1)  < 0.5(1-\alpha)\log\left[ 1+ \frac{P_2} {{(1-\alpha)\sigma_{N_2}}^{2}}\right],
\label{or2}
\end{equation}
\begin{eqnarray}
I (U_1, U_2 ; W_1, W_2) &<&  0.5\alpha\log\left[ 1+ \frac{P_1} {{\alpha\sigma_{N_1}}^{2}}\right]\nonumber\\ 
&+& 0.5(1-\alpha)\log\left[ 1+ \frac{P_2} {{(1-\alpha)\sigma_{N_2}}^{2}}\right].
\label{or3}
\end{eqnarray}

In the following we compare the performance of the AF scheme (explained in sec.~\ref{sec1.6.1}) with the SB scheme. Unlike in the GMAC there is no interference between the two users when orthogonal channels are used. Therefore, in this case we expect AF to perform quite well. 

For AF, the minimum distortions $(D_1,D_2)$ are 
\begin{eqnarray}
D_1&=& \frac{(\sigma_1\sigma_{N_1})^2{\left[P_2(1-{\rho}^2)+\sigma_{N_2}^2\right]}}{P_1P_2(1-\rho^2)+ \sigma_{N_2}^2 P_1+  \sigma_{N_1}^2 P_2+\sigma_{N_1}^2\sigma_{N_2}^2},\label{dist1}\\
D_2&=&\frac{(\sigma_2\sigma_{N_2})^2{\left[P_1(1-{\rho}^2)+\sigma_{N_1}^2\right]}}{P_1P_2(1-\rho^2)+ \sigma_{N_2}^2 P_1+  \sigma_{N_1}^2 P_2+\sigma_{N_1}^2\sigma_{N_2}^2}. 
\label{dist2}
\end{eqnarray}
Thus, as $P_1,P_2 \rightarrow \infty$, $D_1,D_2$ tend to zero. We also see that $D_1$ and $D_2$ are minimum when the average powers used are $P_1$ and $P_2$. These conclusions are in contrast to the case of a GMAC where the distortion for the AF does not approach zero as $P_1,P_2 \rightarrow \infty$ and the optimal powers needed may not be the maximum average allowed $P_1$ and $P_2$ (\cite{Rajesh07allerton}).

We compare the performance of AF with SB for the symmetric case where $P_1=P_2=P, \sigma_1^2=\sigma_2^2=\sigma^2,D_1=D_2=D,\sigma_{N_1}^2=\sigma_{N_2}^2=\sigma_N^2$. These results are from \cite{org}.

We denote the minimum distortions achieved in SB and AF by $D(SB)$ and $D(AF)$ respectively. $\sigma^2$ is taken to be unity without loss of generality. We denote $P/\sigma_N^2$ by $S$. Then

\begin{eqnarray}
D(SB)= \sqrt{\frac{1-\rho^2}{(1+S)^2}+\frac{\rho^2}{(1+S)^4}},\label{distsb}~
D(AF)= \frac{S(1-\rho^2)+1}{1+2S+{S^2}(1-\rho^2)}.\label{distaf}
\end{eqnarray}
We see from the above equations that when $\rho=0,~D(SB)=D(AF)=1/(1+S)$. At high $S,~D(AF) \approx 1/S$ and $D(SB) \approx \sqrt{1-\rho^2}/S$. Eventually both $D(SB)$ and $D(AF)$ tend to zero as $S \rightarrow \infty$. When $S \rightarrow 0$ both $D(SB)$ and $D(AF)$ go to $\sigma^2$.

By squaring the equation \eqref{distsb}  we can show that $ D(AF)\geq D(SB)$ for all $S$. But in \cite{org} we have shown that $D(AF)-D(SB)$ is  small when $S$ is small or large or whenever $\rho$ is small.

\begin{figure}[h]
\centering
\includegraphics [width=2.4in,height=2in] {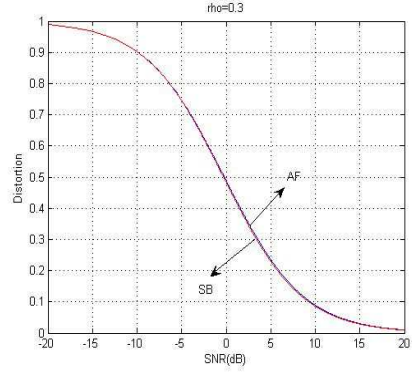}
\caption{SNR vs distortion performance rho=0.3}
\label{fig1.6}
\end{figure}

\begin{figure}[h]
\centering
\includegraphics [width=2.4in,height=2in] {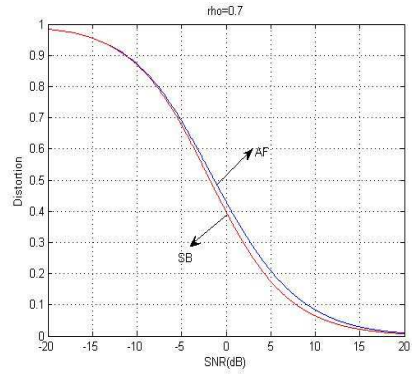}
\caption{SNR vs distortion performance rho=0.7}
\label{fig1.7}
\end{figure}

D(AF) and D(SB) are plotted for $\rho$=0.3 and 0.7 using exact computations in Figs.~\ref{fig1.6} and \ref{fig1.7}.


The above results can be easily extended to the multiple source case. For SB, for the source coding part, the rate region for multiple user case (under a symmetry assumption) is given in \cite{Viswanath}. This can be combined with the capacity achieving Gaussian channel codes over each independent channel to obtain the necessary and sufficient conditions for transmission.

 Let $N$ be the number of sources which are jointly Gaussian with zero mean and covariance matrix $K_U$. Let $P$ be the symmetric power constraint. Let $K_U$ have the same structure as given in \cite{Viswanath}. Let $C_{UY}= \sqrt{P}[1 \rho.....\rho]$ be a $1 \times N$ vector.
The minimum distortion achieved by the AF scheme is given as 
$D(AF)=1-C_{UY} (PK_U+ \sigma_N^2I)^{-1}C_{UY}'$.

\subsection {Side information}
Let us consider the case when side information $Z_i$ is available at encoder $i$, $i=1,2$ and $Z$ is available at the decoder. One use of the side information $Z_i$ at the encoders is to increase the correlation between the sources. This can be optimally done (see~\cite{fried}), if we take appropriate linear combination of $(U_i,Z_i)$ at encoder $i$. The following results are from \cite{org}. We are not aware of any other result on performance of joint source-channel schemes with side information. We are currently working on obtaining similar results for the general MAC. 
\subsubsection{AF with side information}
{\bf{Side information at encoders only}}:
A linear combination of the source outputs and side information $L_i=a_iU_i+b_iZ_i,~i=1,2$ is amplified and sent over the channel. We find   the linear combinations, which minimize the sum of distortions. For this we consider the following optimization problem:\\
Minimize
\begin{equation} 
D(a_1,b_1,a_2,b_2)=E[(U_1-\hat{U}_1)^2]+E[(U_2-\hat{U}_2)^2]
\label{optimi}
\end{equation} 
subject to 
\begin{equation*}
E[X_1^2]\leq P_1 ,~ E[X_2^2]\leq P_2
\end{equation*}
where 
\begin{equation*}
\hat{U}_1=E[U_1|Y_1,Y_2],~\hat{U}_2=E[U_2|Y_1,Y_2].
\end{equation*}

{\bf{Side information at Decoder only}}:
In this case the decoder side information $Z$ is used in estimating $(U_1,U_2)$ from $(Y_1,Y_2)$. The optimal estimation rule is
\begin{equation} 
\hat{U}_1=E[U_1|Y_1,Y_2,Z],\hat{U}_2=E[U_2|Y_1,Y_2,Z].
\label{optim1}
\end{equation} 

{\bf{Side information at both Encoder and Decoder}}:
Linear combinations of the sources are amplified as above and sent over the channel. To find the optimal linear combination, solve an optimization problem similar to \eqref{optimi} with $(\hat{U}_1,\hat{U}_2)$ as given in \eqref{optim1}.

\subsubsection{SB with side information}
For a given $(L_1,L_2)$ we use the source-channel coding scheme explained at the end of sec.~\ref{sec1.7.1}. The side information $Z$ at the decoder reduces the source rate region. This is also used at the decoder in estimating $(\hat{U}_1,\hat{U}_2)$. The linear combinations $L_1$ and $L_2$ are obtained which minimize \eqref{optimi} through this coding-decoding scheme.

\subsubsection{Comparison of AF and SB with side information}
We provide the comparison of AF with SB for $U_1,U_2 \sim \mathcal{N}(0,1)$. Also we take the side information with a specific structure  which seems natural in this set up. Let $Z_1= s_1U_2+V_1$ and $Z_2= s_2U_1+V_2$, where $V_1,V_2 \sim \mathcal{N}(0,1)$ and are independent of each other and independent of the sources, and $s_1$ and $s_2$ are constants that can be interpreted as the side channel SNR. We also take $Z=(Z_1,Z_2)$.

\begin{figure}[!h]
\centering
\includegraphics [width=2.4in,height=2in ] {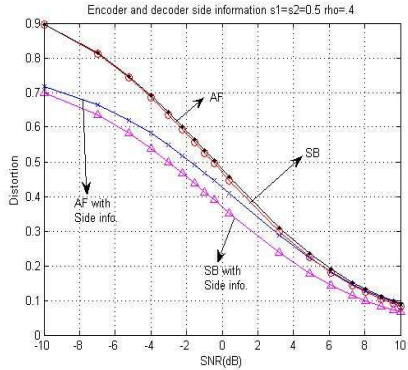}
\caption{AF and SB with both encoder and decoder side information}
\label{last_fig}
\end{figure}

We have compared AF and SB with different $\rho$ and $s_1,s_2$ by explicitly computing the minimum $(D_1+D_2)/2$ achievable. We take $P_1=P_2$. For $s_1=s_2=0.5$ and $\rho=0.4$ we provide the results in Fig.~\ref{last_fig}. From the Figure one sees that without side information, the performance of AF and SB is very close for different SNRs. The difference in their performance increases with side information for moderate values of SNR because the effect of the side information is to effectively increase  the correlation between the sources. Even for these cases at low and high SNRs the performance of AF is close to that of SB. These observations are in conformity with our conclusions in the previous Section . 

Our other conclusions, based on computations not provided here are the following.
For the symmetric case, for SB, encoder-only side information reduces the distortion marginally. This happens because a distortion is incurred for $(U_1,U_2)$ while making the linear combinations $(L_1,L_2)$. For the AF we actually see no improvement and the optimal linear combination has $b_1=b_2=0$. For decoder-only side information the performance is improved for both AF and SB as the side information can be used to obtain better estimates of $(U_1,U_2)$. Adding encoder side information  further improves the performance only marginally for SB; the AF performance is not improved.

In the asymmetric case some of these conclusions may not be valid.

\section{MAC with feedback}\label{sec1.8}
In this section we consider a memoryless MAC with feedback. The channel output $Y_{k-1}$ is available to the encoders at time $k$.

Gaarder and Wolf (\cite{gw}) showed that, unlike in the point to point case, feedback increases the capacity region of a discrete memoryless multiple-access channel . In \cite{covfeed} an achievable region 
\begin{eqnarray}
&R_1 < I(X_1;Y|X_2,U),~ R_2 < I(X_2;Y|X_1,U),\label{cl}\\
&R_1+R_2 < I(X_1,X_2;Y)\nonumber
\end{eqnarray}
where $p(u,x_1,x_2,y)=p(u)p(x_1|u)p(x_2|u)p(y|x_1,x_2)$. It was demonstrated in \cite{fmj} that the same rate region is achievable if there is a feedback link to only one of the transmitters. This achievable region was improved in \cite{bl}.

The achievable region for a MAC, where each node receives possibly different channel feedback, is derived in \cite{car}. The feedback signal in their set-up is correlated but not identical to the signal observed by the receiver. A simpler and larger rate region for the same set-up was obtained in \cite{fmj1}.

Kramer (\cite{gkramer}) used the notion of `directed information' to derive an expression for the capacity region of the MAC with feedback. However, no single letter expressions were obtained.

If the users generate independent sequences, then the capacity region $\mathcal{C}_{fb}$ of the white Gaussian MAC is (\cite{Ozarow84capacity})
\begin{eqnarray}
\label{feed}
\mathcal{C}_{fb}=\bigcup_{0 \leq \rho\leq1} \biggl\{ (R_1,R_2):  R_1 \leq 0.5\log\left[ 1+ \frac{P_1(1- {{\rho}}^{2})} {{\sigma_N}^{2}}\right],\nonumber\\
 R_2 \leq  0.5\log\left[ 1+ \frac{P_2(1- {{\rho}}^{2})} {{\sigma_N}^{2}}\right],\\   			        
R_1+R_2 \leq   0.5\log\left[ 1+ \frac{P_1 + P_2 + {2} {{\rho}}{\sqrt{P_1P_2}}} {{\sigma_N}^{2}} \right]\nonumber
\biggr\}.
\end{eqnarray}
The capacity region for a given $\rho$ in \eqref{feed} is same as in ~\eqref{coo0}-\eqref{coo2} for a channel without feedback but with correlation $\rho$ between channel inputs $(X_1,X_2)$. Thus  the effect of feedback is to allow arbitrary correlation in $(X_1,X_2)$.

An achievable region for a GMAC with noisy feedback is provided in \cite{wigg}. Gaussian MAC with different feedback to different nodes is considered in \cite{send1}. An achievable region based on cooperation among the sources is also given.

  Reference~\cite{wu} obtains an achievable region when non-causal state information is available at both encoders. The authors also provide the capacity region for a Gaussian MAC with additive interference and feedback. It is found that feedback of the output enhances the capacity of the MAC with state. Interference when causally known at the transmitters can be exactly cancelled and hence has no impact on the capacity region of a two user MAC. Thus the capacity region is the same as given in \eqref{feed}. 
  
  In \cite{mer}, it is shown that feedback does not increase the capacity of the Gelfand-Pinsker channel (\cite{gf}) and feedforward does not improve the achievable rate-distortion performance in the Wyner-Ziv system (\cite{wyner}).

 MAC with feedback  and correlated sources  (MACFCS) is studied in \cite{ong,murugan}. This has a MAC with correlated sources and a MAC with feedback  as special cases. Gaussian MACFCS with a total average power constraint is considered in \cite{murugan}. Different achievable rate regions and a capacity outer bound are given for the MACFCS in \cite{ong}. For the first achievable region a decode and forward based strategy is used where the sources first exchange their data, and then cooperate to send the full information to the destination. For two other achievable regions, Slepian-Wolf coding is performed first to remove the correlations among the source data and it is followed by the coding for the MAC with feedback or MAC disregarding the feedback. The authors also show that different coding stategies perform better under different source correlation structures.

The transmission of bivariate Gaussian sources  over a Gaussian MAC with feedback is analyzed in \cite{lap}. The authors show that for the symmetric case, for SNR less than a threshold which is determined by the source correlation, feedback is useless and minimum distortion is achieved by uncoded transmission.

\section{MAC with fading}\label{sec1.9}
A Gaussian MAC with a finite number of fading  states is considered. We provide results when there are $M$ independent sources. The channel state information (CSI)  may be available at the receiver and/or the transmitters. Consider the channel (\cite{proakis})
\begin{equation}
Y_k=\sum_{l=1}^M h_{lk}X_{lk}+N_k
\label{eqnmacf}
\end{equation}
where $X_{lk}$ is the  channel input and $h_{lk}$ is the  fading value at time  $k$ for user $l$. The fading processes $\{h_{lk},~k\geq1\}$  of all users are jointly stationary and ergodic and the stationary distribution has a continuous bounded density. The fading process for the different users are independent. $\{N_k\}$ is the additive white Gaussian noise. All the users are power constrainted to $P$, i.e.,  $E[X_l^2] \leq P$ for all $l$.

Since the source-channel separation holds, we provide the capacity region of this channel.

\subsection{CSI at receiver only}
When the channel fading process $\{h_{lk}\}$ is available at the receiver only, the achievable rate region is the set of rates $(R_1,R_2,...,R_M)$ satisfying

\begin{equation}
\sum_{l\in \mathcal{S}}R_l \leq E \left[log \left(1+\frac{\sum_{l\in \mathcal{S}}\nu_lP}{\sigma^2}\right)\right]
\label{fcsir}
\end{equation}
for all subsets $\mathcal{S}$ of $\{1,2....M\}$, $\nu_l=|h_l|^2$ and $\sigma^2=E[N^2]$. The expectation is over all fading powers $\{\nu_l\},~l\in \mathcal{S}$. One of the performance measures is normalized sum rate per user 
\begin{eqnarray}
\label{fcsir1}
R=\frac{1}{M}\sum_{l=1}^{M}R_l=\frac{1}{M} E\left[ log \left(1+\frac{M P \frac{1}{M}\sum_{l=1}^M\nu_l}{\sigma^2}\right)\right]\nonumber\\
\leq\frac{1}{M} log \left(1+\frac{M P\frac{1}{M}\sum_{l=1}^M E[\nu_l]}{\sigma^2}\right).
\end{eqnarray}

If $E[\nu_l]=1$ for each $l$, then the upper bound equals the capacity of the AWGN channel $\frac{1}{M}log\left[1+\frac{MP}{\sigma^2}\right]$. Also, as $M$ increases, if $\{\nu_l\}$ are $iid$, by Law of Large Numbers (LLN), $R$ will be close to this upper bound. Thus averaging over many users mitigates the effect of fading. This is in contrast to the time /frequency/space averaging.

 The capacity achieving distribution is iid Gaussian for each user and the code for one user is independent of the code for another user (in other words AF is optimal in this case).

\subsection{CSI at both Transmitter and Receiver}
The additional element that is introduced when CSI is provided to the transmitters in addition to  the receiver is dynamic power control which can be done in response to the changing channel state. 

Given a joint fading power $\bold{\nu}=(\nu_1,...,\nu_M)$,  $P_i(\bold{\nu})$  denotes the transmit power allocated to user $i$. Let $P_i$ be the average power constraint for user $i$.
For a given power control policy $\mathcal{P}$
\begin{equation}
\mathcal{C}_f(\mathcal{P})=\biggl\{ {\bf{R}}:{\bf{R}}(\mathcal{S}) \leq E\left[\frac{1}{2} log\left(1+\frac{\sum_{i \in \mathcal{S}}\nu_iP_i(\bold{\nu})}{\sigma^2}\right)\right] for~all~ \mathcal{S} \subset \{1,2,...,M\}\biggr\}
\label{pow}
\end{equation}
denotes the rate region achievable. The capacity region is 
\begin{equation}
\mathcal{C}({\bf{{P}}})=\bigcup_{\mathcal{P} \in \mathcal{F}}\mathcal{C}_f(\mathcal{P})
\label{powe1}
\end{equation}
where $\mathcal{F}$ is the set of feasible power control policies,
\begin{equation}
\mathcal{F}\equiv \{\mathcal{P}: E[P_i(\bold{\nu})]\leq {P}_i,~ for~ i=1,...,M\}.
\end{equation}
Since the capacity region is convex, the above characterization implies that time sharing is not required. 

The explicit characterization of the capacity region exploiting its polymatroid structure is given in \cite{tse}.
For $ P_i=P$ for each $i$ and each $h_i$ having the same distribution, the optimal power control is that only the user with the best channel transmits at a time.  The instantaneous power assigned to the $i_{th}$ user, observing the realization of the fading powers $\nu_1,\nu_2,...,\nu_M$ is
\begin{eqnarray}
P_l(\nu_j,~j=1,...,M)=\begin{cases}
\frac{1}{\lambda} -\frac{1}{\nu_l},~\nu_l > \lambda,~\nu_l>\nu_j~j\ne l\\
0~ otherwise
\end{cases}
\end{eqnarray}
where $\lambda$ is chosen such that the average power constraint is satisfied. This function is actually the well known water filling function (\cite{varia}) optimal  for a single user. This strategy does not depend on the fading statistics but for the constant $ \lambda$. The capacity achieving distribution is Gaussian (thus AF for each user in its assigned slot is optimal).

 Unlike in the single user case the optimal power control may yield substantial gain in capacity. This happens because if $M$ is large, with high probability at least one of the iid fading powers will be large providing a good channel for the respective user at that time instant. 
 
 The optimal strategy is also valid for non equal average powers. The only change being that the fading values are normalized by the Lagrange's coefficients \cite{knopp1}. The extension of this strategy to frequency selective channels is given in \cite{knopp2}.

An explicit characterization of the ergodic capacity region and a simple encoding-decoding scheme for a fading GMAC with common data is given in \cite{nan}. Optimum power allocation schemes are also provided.

\section{Thoughts for practitioners}\label{sec1.10}

Practical schemes for distributed source coding, channel coding and joint source-channel coding for MAC  are of interest. The achievability proofs assume infinite length code words and ignore delay and complexity which make them of limited interest in practical scenarios.

Reference~\cite{zahir} reviews a panorama of practical joint source-channel coding methods for single user systems. The techniques given are hierarchical protection, channel optimized vector quantizers (COVQ), self organizing hypercube (SOH), modulation organized vector quantizer and hierarchichal modulation. 

For lossless distributed source coding, Slepian-Wolf (S-W) (\cite{slepian}) provide the rate region. The underlying idea for construction of practical codes for this system is to exploit the duality between the source and channel coding. The approach is to partition the space of all possible source outcomes into disjoint bins that are cosets of a good linear channel code. Such constructions lead to  constructive and non-asymptotic schemes.

 Wyner was the first to suggest such a scheme  in \cite{wy}. Inspired by Wyner's schme, Turbo/ LDPC based practical code design is given in \cite{liv} for correlated binary sources. The  correlation between the sources  were modelled by a 'virtual' binary symmetric channel (BSC) with crossover probability $p$. The performance of this scheme is very close to the Slepian-Wolf limit $H(p)$. S-W code designs using powerful turbo and LDPC codes for other correlation models and more than two sources is given in \cite{lan}.

 LDPC based codes were also proposed in \cite{eldpc} where a general iterative  S-W decoding algorithm that incorporates the graphical structure of all the encoders and operates in a `Turbo like' fashion is proposed. Reference~\cite{maha} proposes LDPC codes for binary S-W coding problem with Maximum Likelihood(ML) decoding. This gives an upper bound on performance with iterative decoding. They also show that a linear code for S-W source coding can be used to construct a channel code for a MAC with correlated additive white noise.

 In  Distributed Source Coding using Syndromes (DISCUS) (\cite{discus}) Trellis coded Modulation(TCM), Hamming codes and  Reed - Solomon (RS) codes are used for S-W coding. For the Gaussian version of DISCUS, the source is first quantized and then discrete DISCUS is used at both encoder and decoder. 

Source coding with fidelity criterion subject to the availability of side information is addressed in \cite{wyner}. First the source is quantized to the extend allowed by the fidelity requirement. Then S-W coding is used to remove the information at the decoder due to the side information. Since S-W coding is based on channel codes,  Wyner-Ziv coding can be interpreted as a source-channel coding problem. The coding incurres a quantization loss due to source coding and binning loss due to channel coding.  To achieve Wyner-Ziv limit powerful codes need to be employed for both source coding and channel coding.

It was shown  in \cite{zamirlattice} that nested lattice codes can achieve the Wyner-Ziv limit asymptotically, for large dimensions. A practical nested lattice code implemetation is provided in \cite{servetto}. For the BSC correlation model, linear binary block codes are used for lossy Wyner-Ziv coding in \cite{zamirlattice,verdu}.

Lattice codes and Trellis based codes (\cite{forney}) have been used for both source and  channel coding  for the correlated Gaussian sources. A nested lattice construction based on similar sublattices for high correlation is proposed in \cite{servetto}. Another  approach to practical code constructions is based on Slepian-Wolf coded nested quantization (SWC-NQ) which is  a nested scheme followed by binning. Asymptotic performance bounds of SWC-NQ are established in \cite{liu}. A combination of a scalar quantizer and a powerful S-W code is also used for nested  Wyner-Ziv coding. Wyner-Ziv coding based on TCQ and LDPC are provided in \cite{yang}.  A comparison of  different approaches for both Wyner-Ziv coding and classical source coding are provided in \cite{dsc}.

Low density generator matrix (LDGM) codes are proposed for joint source channel coding of correlated sources in \cite{ldgm}. 

Practical code construction for the special case of the CEO problem are provided in \cite{kr,ysxz}.

\section{Directions for future research}\label{sec1.11}
In this report we have provided sufficient conditions for transmission of correlated sources over a MAC with specified distortions. It is of interest to find a single letter characterization of the necessary conditions and to establish the tightness of the sufficient conditions. It is also of interest to extend the above results and coding schemes to sources correlated in time and a MAC with memory. The error exponents are also of interest.

Most of the achievability results in this report use random codes which are inefficient because of large codeword length. It is desirable to obtain power efficient practical codes for side information aware compression that performs very close to the optimal scheme.  

For the fading channels, fairness of the rates provided to different users, the delay experienced by the messages of different users and channel tracking are issues worth pondering. It is also desirable to find the performance of these schemes in terms of scaling behaviour in a network scenario. The combination of joint source-channel coding and network coding is also a new area of research. Another emerging area is the use of joint source-channel codes in MIMO systems and co-operative communication.

\section{Conclusions}\label{sec1.12}
In this report, sufficient conditions are provided for transmission of correlated sources over a multiple access channel. Various previous results on this problem are obtained as special cases.  Suitable examples are given to emphasis the superiority of joint source-channel coding schemes. Important special cases: Correlated discrete sources over a GMAC and Gaussian sources over a GMAC are discussed in more detail. In particular a new joint source-channel coding scheme is presented for discrete sources over a GMAC. Performance of specific joint source-channel coding schemes for Gaussian sources are also compared. Practical schemes like TDMA, FDMA and CDMA are brought into this framework. We also consider a MAC with feedback and a fading MAC. Various practical schemes motivated by joint source-channel coding are also presented. 

\begin{appendix}
\section{Proof of Theorem 1}
The coding scheme involves distributed quantization $(W_1,W_2)$ of the sources and the side information $(U_1,Z_1),(U_2,Z_2)$  followed by a correlation preserving mapping to the channel codewords. The decoding approach involves first decoding  $(W_1,W_2)$  and then obtaining estimate $(\hat{U}_1,\hat{U}_2)$ as a function of $(W_1,W_2)$ and the decoder side information $Z$. We also use the following  Lemmas in the proof.
\begin{lemma}
({\bf{Markov Lemma}}): Suppose $X\leftrightarrow Y \leftrightarrow Z$. If for a given $(x^n,y^n) \in T_\epsilon^n(X,Y)$, $Z^n$ is drawn according to $\prod_{i=1}^np(z_i|y_i)$, then with high probability $(x^n,y^n,Z^n) \in T_\epsilon^n(X,Y,Z)$ for $n$ sufficiently large.
\label {ls}
\end{lemma}

\begin{lemma}
({\bf{Extended Markov Lemma}}): Suppose $W_1\leftrightarrow U_1Z_1 \leftrightarrow U_2W_2Z_2Z$ and  $W_2\leftrightarrow U_2Z_2 \leftrightarrow U_1W_1Z_1Z$. If for a given  $(u_1^n,u_2^n,z_1^n,z_2^n,z^n)$  $\in$  \\  $T_\epsilon^n(U_1,U_2,Z_1,Z_2,Z)$, $W_1^n$ and $W_2^n$ are drawn respectively according to \\ $\prod_{i=1}^np(w_{1i}|u_{1i},z_{1i})$ and  $\prod_{i=1}^np(w_{2i}|u_{2i},z_{2i})$, then with high probability \\ $(u_1^n,u_2^n,z_1^n,z_2^n,z^n,W_1^n,W_2^n)$ $\in$  $T_\epsilon^n(U_1,U_2,Z_1,Z_2,Z,W_1,W_2)$ for $n$ sufficiently large.
\end{lemma}
The proofs of these lemmas are available in \cite{tb} and \cite{thesis} respectively.

We show the achievability of all points in the rate region (1).

$Proof$:  Fix $p(w_1|u_1,z_1), p(w_2|u_2,z_2),p(x_1|w_1),p(x_2|w_2)$ as well as $f_D^{n}(.)$ satisfying the distortion constraints.

$Codebook~ Generation$: Let $R_i^{'}=I(U_i,Z_i;W_i)+\delta, i \in \{1,2\}$ for some $\delta >0$. Generate $2^{nR_i^{'}}$ codewords of length $n$, sampled iid from the marginal distribution $p(w_i), i \in \{1,2\}$. For each $w_i^n$ independently generate sequence $X_i^n$ according to $\prod_{j=1}^n p(x_{ij}|w_{ij}), i \in \{1,2\}$. Call these sequences $x_i(w_i^n), i \in {1,2}$. Reveal the codebooks to the encoders and the decoder.

$Encoding$: For $i \in \{1,2\}$, given the source sequence $U_i^n$ and  $Z_i^n$, the $i^{th}$ encoder looks for a codeword $W_i^n$ such that $(U_i^n,Z_i^n,W_i^n) \in T_{\epsilon}^n(U_i,Z_i,W_i)$ and then transmits $X_i(W_i^n)$ where $T_{\epsilon}^n(.)$ is the set of $\epsilon$-weakly typical sequences (\cite{Cover04elements}) of length $n$.

$Decoding$: Upon receiving $Y^n$, the decoder finds the unique $(W_1^n,W_2^n)$ pair such that $(W_1^n,W_2^n,x_1(W_1^n),x_2(W_2^n),Y^n,Z^n)\in T_{\epsilon}^n$. If it fails to find such a unique pair, the decoder declares an error and incurres a maximum distortion of $d_{max}$.

In the following we show that the probability of error for the encoding decoding scheme tends to zero as $n \rightarrow \infty$. The error can occur because of the following four events {\bf{E1}}-{\bf{E4}}. We show that $P({\bf{Ei}}) \rightarrow 0$, for $i= 1,2,3,4$.

{\bf{E1}} The encoders do not find the codewords. However from rate distortion theory~\cite{Cover04elements}, page 356, $\lim_{n \to \infty}P(E_1)=0$ if $R_i^{'} > I (U_i,Z_i;W_i), i \in {1,2}$.

{\bf{E2}} The codewords are not jointly typical with $Z^n$.
Probobality of this event goes to zero from the extended Markov Lemma (Lemma \ref{ls}).

{\bf{E3}} There exists another codeword $\hat{w}_1^n$ such that $(\hat{w}_1^n,W_2^n,x_1(\hat{w}_1^n),x_2(W_2^n)$,\\ $Y^n,Z^n)\in T_\epsilon^n$.  Define $ \alpha {\buildrel\Delta \over=}$  $(\hat{w}_1^n,W_2^n,x_1(\hat{w}_1^n),x_2(W_2^n),Y^n,Z^n)$. Then,
\begin{eqnarray}
 P({\bf{E3}}) &=& Pr {\{\text{There is} ~ \hat{w}_1^n \ne w_1^n : \alpha \in T_{\epsilon}^n}\}\nonumber \\
 &\leq& \sum_{\hat{w}_1^n \ne W_1^n:(\hat{w}_1^n,W_2^n,Z^n) \in T_{\epsilon} ^n} Pr{\{\alpha \in T_{\epsilon}^n\}}
 \label{e1}
\end{eqnarray}

The  probability term inside the summation in (\ref{e1}) is 

\begin{eqnarray*}
&\leq& \sum_{(x_1(.),x_2(.),y^n):{\alpha \in T_{\epsilon}^n}} Pr\{x_1(\hat{w}_1^n),x_2(w_2^n),y^n|\hat{w}_1^n,w_2^n,z^n\} \\
&=& \sum_{(x_1(.),x_2(.),y^n):{\alpha \in T_{\epsilon}^n}} Pr\{x_1(\hat{w}_1^n)|\hat{w}_1^n\}Pr\{x_2(w_2^n),y^n|w_2^n,z^n\}\\
&\leq& \sum_{(x_1(.),x_2(.),y^n):{\alpha \in T_{\epsilon}^n}} 2^{-n\{H(X_1|W_1)+H(X_2,Y|W_2,Z)-4\epsilon\}}\\
&\leq& 2^{n{H(X_1,X_2,Y|W_1,W_2,Z)}}2^{-n\{H(X_1|W_1)+H(X_2,Y|W_2,Z)-4\epsilon\}} \nonumber.
\label{e2}
\end{eqnarray*}
But from hypothesis, we have
\begin{eqnarray*}
&&H(X_1,X_2,Y|W_1,W_2,Z)- H(X_1|W_1)-H(X_2,Y|W_2,Z) \\
&=&H(X_1|W_1)+H(X_2|W_2)+H(Y|X_1,X_2)-H(X_1|W_1)-H(X_2,Y|W_2,Z)\\
&=&H(Y|X_1,X_2)-H(Y|X_2,W_2,Z)\\
&=&H(Y|X_1,X_2,W_2,Z)-H(Y|X_2,W_2,Z)\\
&=&-I(X_1;Y|X_2,W_2,Z).
\label{e3}
\end{eqnarray*}
Hence,
\begin{gather}
Pr\{(\hat{w}_1^n,W_2^n,x_1(\hat{w}_1^n),x_2(W_2^n),Y^n,Z^n)\in T_{\epsilon}^n\}
\leq 2^{-n\{I(X_1;Y|X_2,W_2,Z)-6\epsilon\}}.
\label{e4}
\end{gather}
Then from \eqref{e1}
\begin{eqnarray}
P({\bf{E3}})&\leq& \sum_{\hat{w}_1^n \ne w_1^n:(\hat{w}_1^n,w_2^n,z^n)\in T_{\epsilon}^n} 2^{-n\{I(X_1;Y|X_2,W_2,Z)-6\epsilon\}}\nonumber\\
 &=& |\{\hat{w}_1^n:(\hat{w}_1^n,w_2^n,z^n)\in T_{\epsilon}^n\}|2^{-n\{I(X_1;Y|X_2,W_2,Z)-6\epsilon\}}\nonumber\\
 &\leq&|\{\hat{w}_1^n\}|Pr\{\hat{w}_1^n,w_2^n,z^n)\in T_{\epsilon}^n\}2^{-n\{I(X_1;Y|X_2,W_2,Z)-6\epsilon\}}\nonumber\\
 &\leq& 2^{n\{I(U_1,Z_1;W_1)+\delta\}}2^{-n\{I(W_1;W_2,Z)-3\epsilon\}}
 2^{-n\{I(X_1;Y|X_2,W_2,Z)-6\epsilon\}}\label{added}\\
 &=&2^{n\{I(U_1,Z_1;W_1|W_2,Z)\}}2^{-n\{I(X_1;Y|X_2,W_2,Z)-9\epsilon-\delta\}}.\nonumber
\end{eqnarray}
The R.H.S of the above inequality tends to zero if $ I(U_1,Z_1;W_1|W_2,Z) < I (X_1;Y|X_2,W_2,Z)$. In \eqref{added} we have used the fact that
\begin{eqnarray*}
I(U_1,Z_1;W_1)&-&I(W_1;W_2,Z)\\
&=&H(W_1|W_2,Z)-H(W_1|U_1,Z_1)\\
&=&H(W_1|W_2,Z)-H(W_1|U_1,Z_1,W_2,Z)\\
&=&I(U_1,Z_1;W_1|W_2,Z).
\label{e6}
\end{eqnarray*}

Similarly, by symmetry of the problem we require 
$ I(U_2,Z_2;W_2|W_1,Z) < I (X_2;Y|X_1,W_1,Z)$.

{\bf{E4}} There exist other codewords $\hat{w}_1^n$ and $\hat{w}_2^n$  such that \\$\alpha {\buildrel\Delta\over =}(\hat{w}_1^n,\hat{w}_2^n,x_1(\hat{w}_1^n),x_2(\hat{w}_2^n),y^n,z^n)\in T_{\epsilon}^n$. Then,
\begin{eqnarray}
 P({\bf{E4}}) &=& Pr {\{\text{There is} ~ (\hat{w}_1^n,\hat{w}_2^n)  \ne (w_1^n,w_2^n) : \alpha \in T_{\epsilon}^n}\}\nonumber \\
 &\leq& \sum_{(\hat{w}_1^n,\hat{w}_2^n) \ne (w_1^n,w_1^n):(\hat{w}_1^n,\hat{w}_2^n,z^n) \in T_{\epsilon}^n} Pr{\{\alpha \in T_{\epsilon}^n\}}.
 \label{e10}
\end{eqnarray}

The  probability term inside the summation in (\ref{e10}) is 

\begin{eqnarray*}
&\leq& \sum_{(x_1(.),x_2(.),y^n):{\alpha \in T_{\epsilon}^n}} Pr\{x_1(\hat{w}_1^n),x_2(w_2^n),y^n|\hat{w}_1^n,\hat{w}_2^n,z^n\} \\
&\leq& \sum_{....}{Pr\{x_1(\hat{w}_1^n)|\hat{w}_1^n\}Pr\{x_2(\hat{w}_2^n)|\hat{w}_2^n\} Pr\{y^n|z^n\}}\\
&\leq& \sum_{(x_1(.),x_2(.),y^n):{\alpha \in T_{\epsilon}^n}} 2^{-n\{H(X_1|W_1)+H(X_2|W_2)+H(Y|Z)-5\epsilon\}}\\
&\leq& 2^{n{H(X_1,X_2,Y|W_1,W_2,Z)}} 2^{-n\{H(X_1|W_1)+H(X_2|W_2)+H(Y|Z)-7\epsilon\}}.\nonumber
\label{e11}
\end{eqnarray*}
But from hypothesis, we have
\begin{eqnarray*}
H(X_1,X_2,Y|W_1,W_2,Z)&-& H(X_1|W_1)-H(X_2|W_2)-H(Y|Z)\\
&=&H(Y|X_1,X_2)-H(Y|Z)\\
&=&H(Y|X_1,X_2,Z)-H(Y|Z)\\
&=&-I(X_1,X_2;Y|Z).
\label{e12}
\end{eqnarray*}
Hence,
\begin{gather}
Pr\{(\hat{w}_1^n,\hat{w}_2^n,x_1(\hat{w}_1^n),x_2(\hat{w}_2^n),y^n,z^n)\in T_{\epsilon}^n\}
\leq 2^{-n\{I(X_1,X_2;Y|Z)-7\epsilon\}}.
\label{e13}
\end{gather}
Then from \eqref{e10}
\begin{eqnarray*}
P({\bf{E4}}) &\leq& \sum_{\substack{(\hat{w}_1^n,\hat{w}_2^n) \ne (w_1^n,w_1^n):\\(\hat{w}_1^n,\hat{w}_2^n,z^n) \in T_{\epsilon}^n}} 2^{-n\{I(X_1,X_2;Y|Z)-7\epsilon\}}\\
 &=& |\{(\hat{w}_1^n,\hat{w}_2^n):(\hat{w}_1^n,\hat{w}_2^n,z^n)\in T_{\epsilon}^n\}|2^{-n\{I(X_1,X_2;Y|Z)-7\epsilon\}}\\
 &\leq&|\{\hat{w}_1^n\}||\{\hat{w}_2^n\}|Pr\{(\hat{w}_1^n,\hat{w}_2^n,z^n)\in T_{\epsilon}^n\}
 2^{-n\{I(X_1,X_2;Y|Z)-7\epsilon\}}\nonumber\\
&\leq& 2^{n\{I(U_1,Z_1;W_1)+I(U_2,Z_2;W_2)+2\delta\}}\\
&&2^{-n\{I(W_1;W_2,Z)+I(W_2;W_1,Z)+I(W_1;W_2|Z)-4\epsilon\}}
 2^{-n\{I(X_1,X_2;Y|Z)-7\epsilon\}}\nonumber\\
 &=&2^{n\{I(U_1,U_2,Z_1,Z_2;W_1,W_2|Z)\}}2^{-n\{I(X_1,X_2;Y|Z)-11\epsilon-2\delta\}}.
\label{e15}
\end{eqnarray*}
The RHS of the above inequality tends to zero if $ I(U_1,U_2,Z_1,Z_2;W_1W_2|Z) < I(X_1,X_2;Y|Z)$.

Thus as $n \rightarrow \infty$, with probability tending to 1, the decoder finds the correct sequence $(W_1^n,W_2^n)$ which is jointly  weakly $\epsilon$-typical with $(U_1^n,U_2^n,Z^n)$.

The fact that $(W_1^n,W_2^n)$ are weakly $\epsilon$-typical with $(U_1^n,U_2^n,Z^n)$ does not guarantee that $f_D^n(W_1^n,W_2^n,Z^n)$ will satisfy the distortions $D_1,D_2$. For this, one needs that $(W_1^n,W_2^n)$ are distortion-$\epsilon$-weakly typical (\cite{Cover04elements}) with   $(U_1^n,U_2^n,Z^n)$. Let $T_{D,\epsilon}^n$  denote the set of distortion typical sequences (\cite{Cover04elements}). Then by strong law of large numbers  $P(T_{D,\epsilon}^n|T_\epsilon^n)\rightarrow 1$ as $n\rightarrow \infty$. Thus the distortion constraints are also satisfied by $(W_1^n,W_2^n)$ obtained above with a probability tending to 1 as $n \rightarrow \infty$. Therefore, if distortion measure $d$ is bounded 
$\lim_{n \rightarrow \infty}E[d(U_i^n,\hat{U}_i^n)] \leq D_i+\epsilon~i=1,2$.

If there exist $u_i^*$ such that $E[d_i(U_i,u_i^*)]<\infty,~i = 1,2 $, then the result extends to unbounded distortion measures also as follows. Whenever the decoded $(W_1^n,W_2^n)$ are not in the distortion typical set then we estimate $(\hat{U}_1^n,\hat{U}_2^n)$ as $({u_1^*}^n,{u_2^*}^n)$. Then for $i=1,2$,
\begin{equation}
\label{ala}
E[d_i(U_i^n,\hat{U}_i^n)] \leq D_i+\epsilon + E[d(U_i^n,{u_i^*}^n) {\bf{1}}_{\{(T_{D,\epsilon}^n)^c\}}].
\end{equation}
Since $E[d(U_i^n,{u_i^*}^n)] < \infty $ and $P[({T_{D,\epsilon}^n})^c] \rightarrow 0$ as $n \rightarrow \infty$, the last term of \eqref{ala} goes to zero as  $n \rightarrow \infty$.
\end{appendix}
\bibliographystyle{abbrv}
\bibliography{mybibfile}

\end{document}